\documentclass[aip,jcp,reprint]{revtex4-1}
\usepackage[utf8x]{inputenc}
\usepackage[english]{babel}
\usepackage{blindtext}
\usepackage{amsmath,amsthm,amssymb}

\usepackage{epsfig}
\usepackage{graphicx}
\usepackage{subfig}

\usepackage[vlined, ruled]{algorithm2e}

\usepackage{dcolumn}

\usepackage{hyperref}

\begin{document}
\title{Efficient Bayesian estimation of Markov model transition matrices with given stationary distribution}

\author{Benjamin Trendelkamp-Schroer}
\email{benjamin.trendelkamp-schroer@fu-berlin.de}
\author{Frank No\'{e}}
\email{frank.noe@fu-berlin.de}
\thanks{``corresponding author''}

\affiliation{Institut f\"{u}r Mathematik und Informatik, FU Berlin, Arnimallee 6, 14195 Berlin}

\date{\today}

\begin{abstract}
Direct simulation of biomolecular dynamics in thermal equilibrium is challenging due to the metastable nature of conformation dynamics and the computational cost of molecular dynamics. Biased or enhanced sampling methods may improve the convergence of expectation values of equilibrium probabilities and expectation values of stationary quantities significantly. Unfortunately the convergence of dynamic observables such as correlation functions or timescales of conformational transitions relies on direct equilibrium simulations. Markov state models are well suited to describe both, stationary properties and properties of slow dynamical processes of a molecular system, in terms of a transition matrix for a jump process on a suitable discretization of continuous conformation space. Here, we introduce statistical estimation methods that allow a priori knowledge of equilibrium probabilities to be incorporated into the estimation of dynamical observables. Both, maximum likelihood methods and an improved Monte Carlo sampling method for reversible transition matrices with fixed stationary distribution are given. The sampling approach is applied to a toy example as well as to simulations of the MR121-GSGS-W peptide, and is demonstrated to converge much more rapidly than a previous approach in \cite{noe2008}.
\end{abstract}


\maketitle
\section{Introduction}
Characterization of the conformational dynamics of proteins and other biomolecules in thermal equilibrium includes the identification of their metastable
states, and quantification of their populations and transition rates. Such a characterization is essential to analyze and potentially manipulate biologically important conformational transitions, including folding, ligand binding, and aggregation. Unfortunately, a direct observation of dynamical processes with an atomistic resolution is impossible because the scale of conformation dynamics lies well below the diffraction limit of optical methods. Spectroscopic methods that provide information in atomistic detail such as X-ray crystallography do usually only provide information about static quantities. NMR spectroscopy methods provide only indirect observations of dynamical processes via relaxation dispersion correlations whose interpretation is challenging and do not provide direct structural information. Single-molecule spectroscopic methods can probe the dynamical fluctuations of one to two observables directly, but they do not reveal molecular structures.

The recent increase in computing power has enabled the study of conformation dynamics in atomistic detail via direct molecular dynamics
simulations \cite{noe2009,shaw2010,voelz2010,bowman2011helix,kresten2011,sadiq2012}. Nonetheless, the metastable nature of conformation dynamics \cite{elber1987,honeycutt1990,nienhaus1992,schuette2003,noe2007} in combination with the necessary explicit treatment of fast degrees of freedom in the numerical integration of the equations of motions renders the spontaneous observation of rare events on the milliseconds timescales or slower difficult. As a result, one faces severe difficulties when trying to converge expectation values of observables depending on slow processes, such as the implied time scales of large scale conformational changes \cite{clarage1995}. 

The recent years have seen the development of a host of biased or enhanced sampling methods to accelerate rare events, and thus to permit
the efficient exploration of the system's relevant conformations and estimation of at least its thermodynamic quantities, such as the stationary 
probabilities of states and stationary expectation values. To name only some of the best-known examples, replica exchange or parallel tempering methods facilitate the hopping over energetic barriers by exchanging molecular conformations between simulations at different temperatures \cite{sugita1999,rao2003}. 
Flooding methods obtain stationary probabilities by filling up the free energy landscape according to the frequency of visits by the evolving trajectory \cite{grubmueller1995, laio2002}. Umbrella sampling \cite{torrie1977} proceeds by choosing an appropriate re-weighting function restricting the chain to a subspace relevant to the estimation of a chosen observable. An improved version using the weighted histogram analysis method \cite{ferrenberg1989} guides the simulation along a multidimensional hyper-surface specified by a set of a priori chosen reaction coordinates \cite{kumar1995}. For a short pedagogical overview of enhanced ensemble methods see \cite{trebst2006}. Applications include replica exchange folding studies of a Small RNA hairpin \cite{garcia2008}, single-copy tempering for trpzip2, trp-cage, and the villin headpiece \cite{zhang2010}, as well as reconnaissance meta-dynamics for the binding of benzamidine to trypsin \cite{soderhjelm2012}. Examples for problems that have also been successfully treated are first and second order phase transitions in lattice spin systems \cite{wang2001}.

While biased or enhanced sampling methods can generate estimates of equilibrium quantities efficiently, they usually do not preserve the equilibrium dynamics. Thus, dynamical observables such as rates or time-correlation functions have to be estimated using other methods, chiefly from direct equilibrium molecular dynamics simulations.
An approach frequently used to integrate and analyze molecular dynamics data is Markov modeling \cite{schuette1999, swope2004, singhal2004, schultheis2005, chodera2007, noe2007, pan2008, prinz2011}. Markov models approximate the continuous phase space dynamics in terms of a discrete space Markov jump process. A particular advantage of this approach is that Markov processes have been extensively studied in Mathematics so that there are a large number of rigorous results available. The construction of Markov models proceeds through first choosing a suitable discretization of conformation space and then estimating a transition probability matrix from counted transitions between conformational subsets specified by the discretization \cite{schuette1999}. Choosing the discretization so as to achieve an accurate Markov model is a topic of current research \cite{chodera2007, sarich2010, chodera2011,perez2012}. As shown in \cite{sarich2010} the approximation error can be bounded and vanishes as the discretization gets finer and the lag time is increased. A recently outlined variational method \cite{nueske2012} can be employed to approximate relevant spectral properties of the transition operator by an application of the famous Rayleigh-Ritz principle. An approach using basis functions and variational inequalities makes it possible to connect to established methods from electronic structure calculations and may proof useful in iteratively improving conformation space discretization. For an overview of the Markov state model approach to conformation dynamics see \cite{prinz2011}. The Markov state model approach has been able to reconstruct complex molecular processes such as protein folding \cite{schuette1999, swope2004, singhal2004, schultheis2005, chodera2007, noe2007, pan2008, noe2009, prinz2011, lane2011, bowman2011}, natively unstructured protein dynamics \cite{perez2012}, and protein-ligand binding \cite{beauchamp2011, held2011, BuchFabritiis_PNAS11_Binding, HuangCaflisch_PlosCB11_SmallMoleculeUnbinding, bowman2012} from computer generated trajectories. In addition the Markov model framework allows the comparison of simulation driven predictions with experimental findings in a consistent manner \cite{SzerRoux_JCP08_MSM-scattering, Zhuang_JPCB11_MSM-IR, noe2011, keller2012}.

Since enhanced and biased sampling methods can significantly improve the convergence of stationary quantities in the presence of long timescales, 
while direct molecular dynamics simulations can probe dynamical quantities depending on short timescales, it would be desirable to combine the
advantages of both approaches. A natural mathematical basis to foster this combination is detailed balance of the dynamics. Detailed balance states that
under equilibrium conditions, the ratio of stationary probabilities between two states is equal to the inverse ratio of transition rates or probabilities
between them. On the microscopic scale, detailed balance is a natural consequence of the time inversion invariance of the microscopic equations of motion and the Gaussian white noise nature of the stochastic fluctuations \cite[p. 88ff.]{lelievre2010}. When using Markov models, microscopic detailed balanced directly translates into detailed balance on the level of Markov states. Therefore, it would be desirable to include prior information of the stationary distribution into the estimation of dynamical observables such as correlation functions and time scales or rates of conformational changes. One could for example use well converged equilibrium probabilities estimated on conformational subsets constituting a suitable discretization from an extended ensemble simulation and generate observations of equilibrium fluctuations from a standard equilibrium simulation. The precise knowledge of the stationary probabilities could for example be used to obtain sharper estimates of dynamical quantities such as timescales for large scale conformational transitions. 

Detailed balance is now commonly used as a constraint to guide the maximum likelihood estimation of Markov model transition matrices from
observed transition counts \cite{Bowman_JCP09_Villin, prinz2011} However, these existing approaches do not permit to explicitly include prior knowledge of the stationary distribution. Beyond maximum likelihood estimates, the estimation of statistical uncertainty stemming from the fact that only finitely many transition counts have
been observed, is crucial to allow a meaningful comparison with expectation values obtained from other simulations as well as with observations from experiments to be made \cite{chodera2010}. Furthermore, quantification of statistical uncertainties is a prerequisite to guide an adaptive sampling
approach that aims at reducing them efficiently \cite{singhal2005,hinrichs2007,Bowman_JCP09_Villin}.
In Singhal et al. \cite{singhal2005} direct sampling of transition matrices was applied to calculate the distribution of mean first passage times. A computationally efficient procedure to estimate the variance together with the mean based on a Gaussian approximation of the distribution of transition matrices and a first order Taylor expansion of the target observable was also developed. In \cite{hinrichs2007} the method was extended to the estimation of eigenvalues and eigenvectors. In \cite{prinzHeld2011}, a similar perturbation method was used to evaluate the statistical error of committor probabilities. In \cite{roeblitz2008} a related approach based on perturbation theory of spectral subspaces is developed in order to achieve a refinement of a grid-free conformation space discretization. 
A full Bayesian approach for estimating statistical errors including the detailed balance constraint was introduced in \cite{noe2008}. In a subsequent
study, we have extended the formalism by also including statistical uncertainties of spectroscopic observables \cite{chodera2010}. 
Ref. \cite{bacallado2009} has used a different approach, an edge reinforced random walk, to sample reversible transition matrices.
As yet, the Markov chain Monte Carlo approach in Ref. \cite{noe2008} is the only approach that permits to explicitly include prior knowledge of the stationary distribution into the estimation of the probability distribution of transition matrices. However, this sampler has rather poor mixing properties, thus requiring many iterations and a high computational load before the probability distributions can be estimated reliably.

In the following we will introduce efficient methods to include prior knowledge of the stationary distribution into reversible transition matrix estimates:
(1) Maximum likelihood estimation methods are given that either solve a constrained convex optimization problem using standard optimization libraries, or proceed via an iterative likelihood maximization algorithm.
(2) An efficient Gibbs method is introduced to sample the conditional densities of individual transition matrix elements, offering improved convergence properties over the previous approach in Ref. \cite{noe2008}. 
The estimation and sampling methods described here are implemented in the EMMA Markov model toolkit \cite{senne2012}. The maximum likelihood estimation for fixed stationary distribution can be performed using the EMMA command \texttt{mm\_estimate} and the Gibbs sampling of reversible transition matrices with fixed stationary distribution is available via the command \texttt{mm\_transitionMatrixSampling}.

\section{Probability distributions for transition matrices}
If one has at hand only a finite observation $X_1,\dotsc,X_N$ of a Markov jump process there are usually an infinite number of transition matrices $P$ that are compatible with the given data. In the following we assume that one can directly observe transitions between individual micro states $i \in {1,\dotsc,n}$. A single entry $p_{ij}$ of a transition matrix quantifies the probability to make a transition to state $j$ given that you have started in $i$,\[p_{ij}=\mathbb{P}(X_{k+1}=j|X_{k}=i).\] If the micro state jump process is Markovian the probability of observing a certain realization of the process $X_1,\dotsc,X_N$ depends only on the number of transitions between pairs of states in $X_1,\dotsc,X_N$ together with the probability to start in $X_1$. Thus the matrix of transition counts $C$ together with the probability of the initial state, $p(X_1)$, completely determines the probability of a given observation for a fixed $P$, \begin{equation}\label{eqn:partial_exchangeability} p(X_1,\dotsc,X_N|P)=p(C|P)p(X_1).\end{equation} As a result of Markovianity the probability of observing transition counts $c_{ij}$ given a set of transition probabilities $p_{ij}$ is given by the multinomial distribution  \begin{equation}\label{eqn:probability_observation} p(C|P)\propto \prod_{i,j=1}^{n} p_{ij}^{c_{ij}}.\end{equation} However, we need the probability of a certain transition matrix given an observation of transition counts, $p(P|C)$. Bayes' theorem can be used to relate $p(C|P)$ to $p(P|C)$ via \[p(P|C) \propto p(C|P)p(P).\] Using a suitable conjugate prior with prior counts $b_{ij}$, as outlined in \cite{prinz2011}, we find that this probability is given by a product of Dirichlet distributions \begin{equation}\label{eqn:probability_transition_matrix} p(P|C) \propto \prod_{i,j=1}^{n} p_{ij}^{c_{ij}+b_{ij}}.\end{equation} Here the following normalization condition for row-stochasticity of $P$ is assumed to hold, \begin{align}\label{eqn:normalization} &\sum_{k=1}^{n}p_{ik}=1  & i=1,\dotsc,n.\end{align} The structure of \eqref{eqn:probability_transition_matrix} makes it possible to generate independent Dirichlet distributed rows if no additional constraints on $P$ are imposed \cite{singhal2005, hinrichs2007, kube2008, devroye1986}. If one desires to restrict the space of all admissible transition matrices to those obeying a detailed balance condition \begin{equation} \label{eqn:detailed_balance} \pi_i p_{ij}=\pi_j p_{ji} \end{equation} the additional interdependence between rows prohibits to generate samples from \eqref{eqn:probability_transition_matrix} by direct sampling of individual rows. In \cite{noe2008} a Metropolis Hastings Monte Carlo chain method is developed to generate random transition matrices from \eqref{eqn:probability_transition_matrix} under the detailed balance constraint. In the following we will only consider the situation in which the stationary probabilities have been already computed using a different simulation algorithm. Note that fixing $\pi_1,\dotsc,\pi_n$ and requiring detailed balance reduces the number of independent variables $p_{ij}$ from $n(n-1)$ to $\frac{n(n-1)}{2}$. This is a 50\% reduction in dimension and we expect that imposing this extra symmetry will have a large effect when comparing quantities estimated with and without these constraints. In the following we will use the normalization condition \eqref{eqn:normalization} to determine the diagonal of $P$ from the off-diagonal elements, \begin{align*}& p_{ii}=1-\sum_{k\neq i} p_{ik} & i=1,\dotsc,n \end{align*} and the detailed balance condition \eqref{eqn:detailed_balance} in combination with the fixed stationary vector to determine the lower triangular part of $P$ from the upper triangular one, \begin{align*} & p_{ji} = \frac{\pi_i}{\pi_j} p_{ij} & 1\leq i<j \leq n.\end{align*}

This approach for incorporating a priori knowledge about stationary probabilities has a straightforward generalization to situations in which the stationary probabilities are not precisely known. If one has obtained a probability model for the stationary probabilities \begin{equation}\label{eqn:probability_stationary_distribution} p(\pi|E)\end{equation} from an enhanced sampling method $E$ one can incorporate this prior knowledge of $\pi$ into a probability model for $P$. The probability model for $P$ given the evidence $C$ and $E$ is given by \begin{equation}\label{eqn:probability_stationary_and_transitions} p(P|C,E)=\int \mathrm{d}\pi \, p(P|C,\pi)p(\pi|E).\end{equation} $P\sim p(P|C,E)$ can be sampled by iteratively generating samples of $\pi$ from $p(\pi|E)$ and of $P$ from $p(P|C,\pi)$. 

\section{Conditional probabilities \label{sec:conditional_probabilities}}
The Gibbs sampling strategy facilitates sampling of a joint distribution by generating random variates from the conditionals. In the following we will show that for a fixed stationary vector $\left(\pi_1,\dotsc,\pi_n \right)$ all the conditionals of $p(P|C)$ have a simple analytical form. Furthermore we will outline a method to generate random variates efficiently from all conditionals for all possible configurations of $P$ and $\pi$. For the sake of brevity of notation we will often supress the fixed observation $C$ when stating relations for the conditionals. There are only four factors in the joint probability \eqref{eqn:probability_transition_matrix} with an explicit dependence on the transition matrix element $p_{ij}$. The element $p_{ii}$ is linked to $p_{ij}$ by constraint \eqref{eqn:normalization}, $p_{ji}$ is related to $p_{ij}$ by \eqref{eqn:detailed_balance}, and finally $p_{jj}$ is dependent on $p_{ij}$ by a combination of \eqref{eqn:normalization} and \eqref{eqn:detailed_balance}. For this reason the conditional probability for $p_{ij}$ is given conditioned on the following set of transition matrix elements \[\{p_{11},\dotsc,p_{nn}\}/\{p_{ii}, p_{ij}, p_{ji}, p_{jj}\}.\] In a slight abuse of notation we indicate this conditioning on the above set writing the conditional density for $p_{ij}$ as $p(p_{ij}|p_{k \neq i,j, l\neq i,j})$. It is given by \[p(p_{ij}|p_{k\neq i,j, l\neq i,j})\propto p_{ij}^{c_{ij}}p_{ji}^{c_{ji}}p_{ii}^{c_{ii}}p_{jj}^{c_{jj}}.\] Plugging in the constraints \eqref{eqn:normalization}, \eqref{eqn:detailed_balance} we get \begin{align*} p(p_{ij}|p_{k\neq i,j,l\neq i,j}) \propto & p_{ij}^{c_{ij}+c_{ji}} \\ & \times ((1-\sum_{k\neq i,j}p_{ik})-p_{ij})^{c_{ii}} \\  & \times ((1-\sum_{k\neq j,i}p_{jk})-\frac{\pi_i}{\pi_j}p_{ij})^{c_{jj}} \end{align*} explicitly showing the unvariate dependence on $p_{ij}$. Now we define \begin{equation}\label{eqn:def_delta}\Delta_{ij}=(1-\sum_{k\neq i,j}p_{ik}),\end{equation}\begin{equation}\label{eqn:def_gamma}\Lambda_{ij}=\frac{\pi_j}{\pi_i}(1-\sum_{k\neq j,i}p_{jk}).\end{equation} Using these we can rewrite the conditional density as \begin{equation}\label{eqn:conditional_density_raw} p(p_{ij}|p_{k\neq i,j,l\neq i.j})\propto p_{ij}^{c_{ij}+c_{ji}}(\Delta_{ij} - p_{ij})^{c_{ii}}(\Lambda_{ij}-p_{ij})^{c_{jj}}.\end{equation} We assume that $\Delta_{ij} \leq \Lambda_{ij}$. Then we can define \[x=\frac{p_{ij}}{\Delta_{ij}}\] and define the following parameters, \begin{align} & a=c_{ij}+c_{ji}, \\ & b=c_{ii}, \\ & c={c_{jj}}, \\ & d=\frac{\Lambda_{ij}}{\Delta_{ij}}.\end{align} In the case $\Delta_{ij}>\Lambda_{ij}$ we switch the definition of $b$, $c$, define $d=\Delta_{ij}/\Lambda_{ij}$ and $x=p_{ij}/\Lambda_{ij}$. It can be seen that in both cases $a,b,c \geq 0$, $d\geq 1$, and $0\leq x \leq 1$. After a little algebra we get \begin{equation}\label{eqn:conditional_density} p(x|a,b,c,d)\propto x^a (1-x)^b (d-x)^c \end{equation} with $0\leq x \leq 1$. This means that if we can generate random variates from $p(x|a,b,c,d)$ efficiently for all admissible parameters, we can efficiently sample all conditional densities arising during a Gibbs sampling procedure. The dependence of the conditionals for $p_{ij}$ on both $c_{ij}$ and $c_{ji}$ clearly reflects the additional symmetry imposed by the detailed balance condition.

\subsection{Log-concave densities}
We can write the density $p(x|a,b,c,d)$ in the following way, \[p(x|a,b,c,d)=e^{q(x|a,b,c,d)},\] with \[q(x|a,b,c,d)=a\log(x)+b\log(1-x)+c\log(d-x).\] The second derivative of $q(x|a,b,c,d)$ is given by \[q''(x|a,b,c,d)=-\frac{a}{x^2}-\frac{b}{(1-x)^2}-\frac{c}{(d-x)^2}.\] It is easy to see that \[q''(x|a,b,c,d) \leq 0\] for all $0\leq x \leq 1$ and all parameters $a,b,c \geq 0$ and $d \geq 1$. This is a sufficient condition for $q(x|a,b,c,d)$ to be a concave function and therefore all conditionals $p(x|a,b,c,d)$ fall into the category of log-concave densities. There exist efficient approaches for the generation of random variates from a log-concave density given explicit knowledge of the mode point and the ability to evaluate the density $p(x)$ and the first derivative of its logarithm $q(x)=\log p(x)$. For an overview of methods to sample from log-concave densities see \cite{devroye1986}. The crucial feature employed by all these methods is that any concave function $q: \Omega \rightarrow \mathbb{R}$ is bounded from above by all its tangents, so that for all $x_0$ for which $q'(x_0)$ exists, the following holds \begin{align*} & q(x) \leq q(x_0)+q'(x_0)(x-x_0) & \forall x \in \Omega. \end{align*} Since the exponential function is a monotone function we have \[f(x)=e^{q(x)}\leq e^{q(x_0)+q'(x_0)(x-x_0)}.\] The global maximum or mode point of $p(x|a,b,c,d)$ is attained at $x_m$ with \[q'(x_m|a,b,c,d)=0\] subject to the constraint $0\leq x_m \leq 1$. We have \begin{align*}q'(x|a,b,c,d)= & x^{-1}(1-x)^{-1}(d-x)^{-1} \\ & \left\{a(1-x)(d-x)-bx(d-x)-cx(1-x)\right\}.\end{align*} It is obvious that it suffices to find the zeros of \[a(1-x)(d-x)-bx(d-x)-cx(1-x).\] This expression is at most quadratic in $x$ for all admissible parameters. Therefore extremal points of $q(x|a,b,c,d)$ are given by \[x_{1,2}=\frac{1}{2(a+b+c)}\left((a+b)d+(a+c)\pm\sqrt{r}\right), \] with \[r=\left[(a+b)d+(a+c) \right]^2 -4(a+b+c)ad.\] It is apparent that $p(x|a,b,c,d)$ has zeros at $x_0=0$, $x_0=1$ and $x_0=d$. Recall that $d \geq 1$. This means that there is one extremal point in $[0,1]$ and one extremal point in $[1,d]$. Therefore we conclude that $x_m$ corresponds to the smaller one of the two extremal points, \begin{equation}\label{eqn:mode_point}x_m=\frac{1}{2(a+b+c)}\left((a+b)d+(a+c)-\sqrt{r}\right). \end{equation} We note that the mode point need not lie in the interior of the unit interval so that $x_m=0$ and $x_m=1$ are possible values.

\subsection{Optimal piecewise approximation}
\begin{figure}[htb]
 \includegraphics[width=0.45\textwidth]{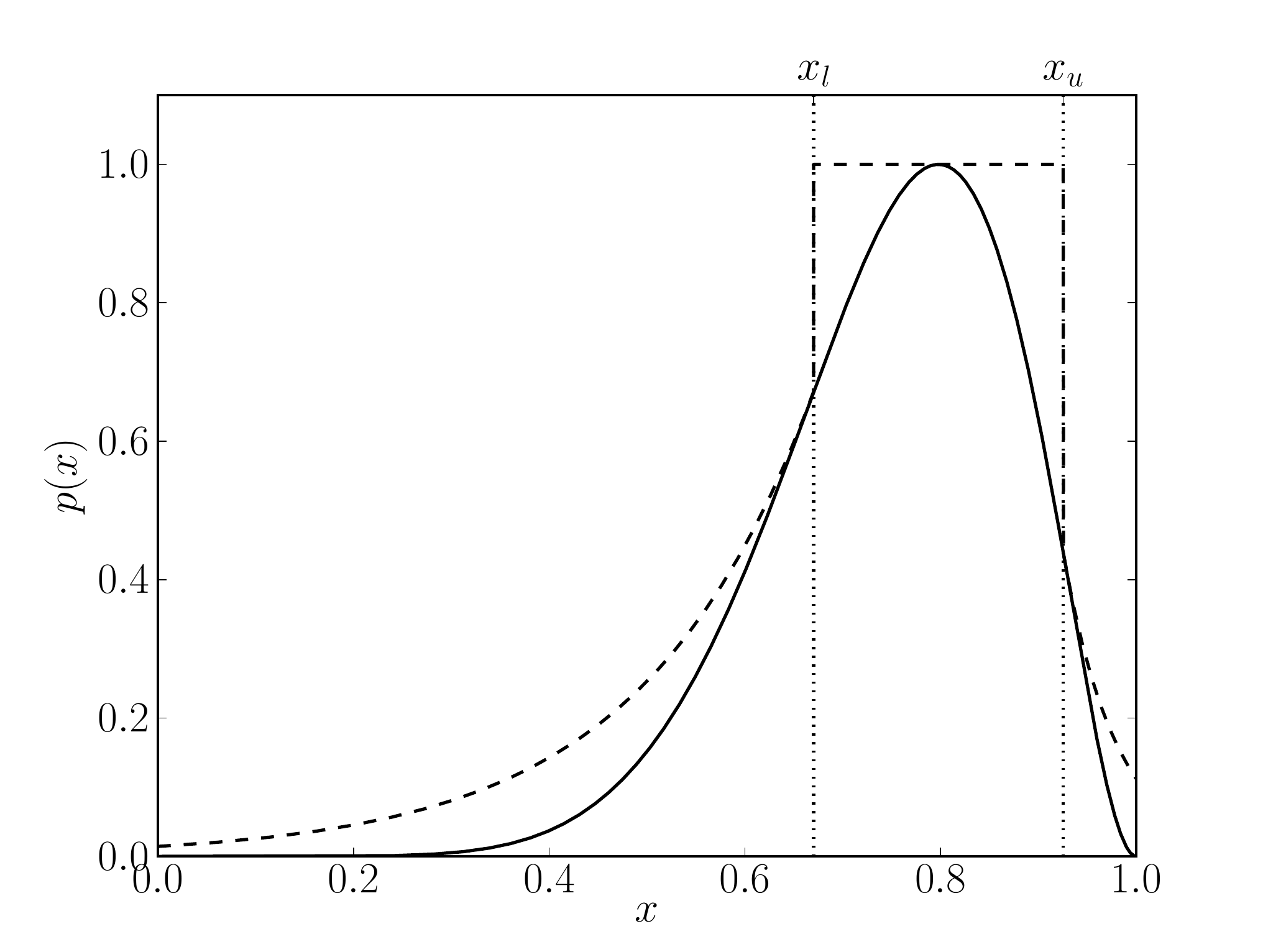}
  \caption{The conditional density $p(x|a,b,c,d)$ for $a=8.0$, $b=2.0$, $c=4.0$ and $d=30.0$ (solid line) and the corresponding enveloping function $g(x)$ (dashed line). The density was scaled to the mode point value  $p(x_m|a,b,c,d)$ to fit it into the range $[0,1]$.}
\label{fig:enveloping_density}
\end{figure}

We will use a piecewise enveloping function $g(x|a,b,c,d)$ bounding $p(x|a,b,c,d)$ consisting of a uniform density around the mode point and exponential tails elsewhere. For log concave densities $f(x)$ it is possible to use the following general approach to find an enveloping function $g(x)$ for $f(x)$. Let again $q(x)=\log f(x)$. Consider the following piecewise defined function $h(x)$, \[h(x)=
\begin{cases}
q(x_l)+q'(x_l)(x-x_l) & -\infty < x < x_l \\
q(x_m)  & x_l \leq x \leq x_u \\
q(x_u)+q'(x_u)(x-x_u) & x_u \leq x < \infty        
\end{cases}.
\] Here $x_l$ and $x_u$ denote the lower and the upper bound for a region around $x_m$ in which $f(x)$ will be bounded by a uniform density $f(x_m)\chi_{[x_l,x_u]}(x)$. As a consequence of concavity the function $h(x)$ is a valid dominating function for $q(x)$. Thus $g(x)=e^{h(x)}$ is a valid enveloping function for $f(x)$. \autoref{fig:enveloping_density} shows $p(x)$ and the enveloping density $g(x)$.
The optimal choice for $x_l$ and $x_u$ is the one that minimizes the area between $g(x)$ and $f(x)$ leading to the lowest possible rejection rate. One can show \cite{devroye1986} that an $x_l \leq x_m$ and $x_u \geq x_m$ is optimal if \begin{align*} & f(x_l^{*})=\frac{f(x_m)}{e}, & f(x_u^{*})=\frac{f(x_m)}{e}.\end{align*} Here $e$ denotes the Euler number. If $f^{-1}$ is explicitly known finding the optimal solution is straightforward. It is apparent that for unimodal (continuous) densities $f$ the inverse $f^{-1}$ is always unique on $x \leq x_m$ and on $x \geq x_m$. Unfortunately we do not have an explicit expression for $p^{-1}(x|a,b,c,d)$. It is however always possible to choose suboptimal points $x_l$ and $x_u$ at the cost of a larger rejection rate. In the case that $x_m$ lies on the boundary of the domain of $f(x)$ the bounding function will be a single exponential function. In the case that $x_l$ or $x_u$ lie outside the domain of definition one restricts $h(x)$ by truncating it a the boundary points so that only one exponential tail or only the constant part will survive.

\subsection{Suboptimal piecewise approximation}
\begin{figure}[htb]
 \includegraphics[width=0.45\textwidth]{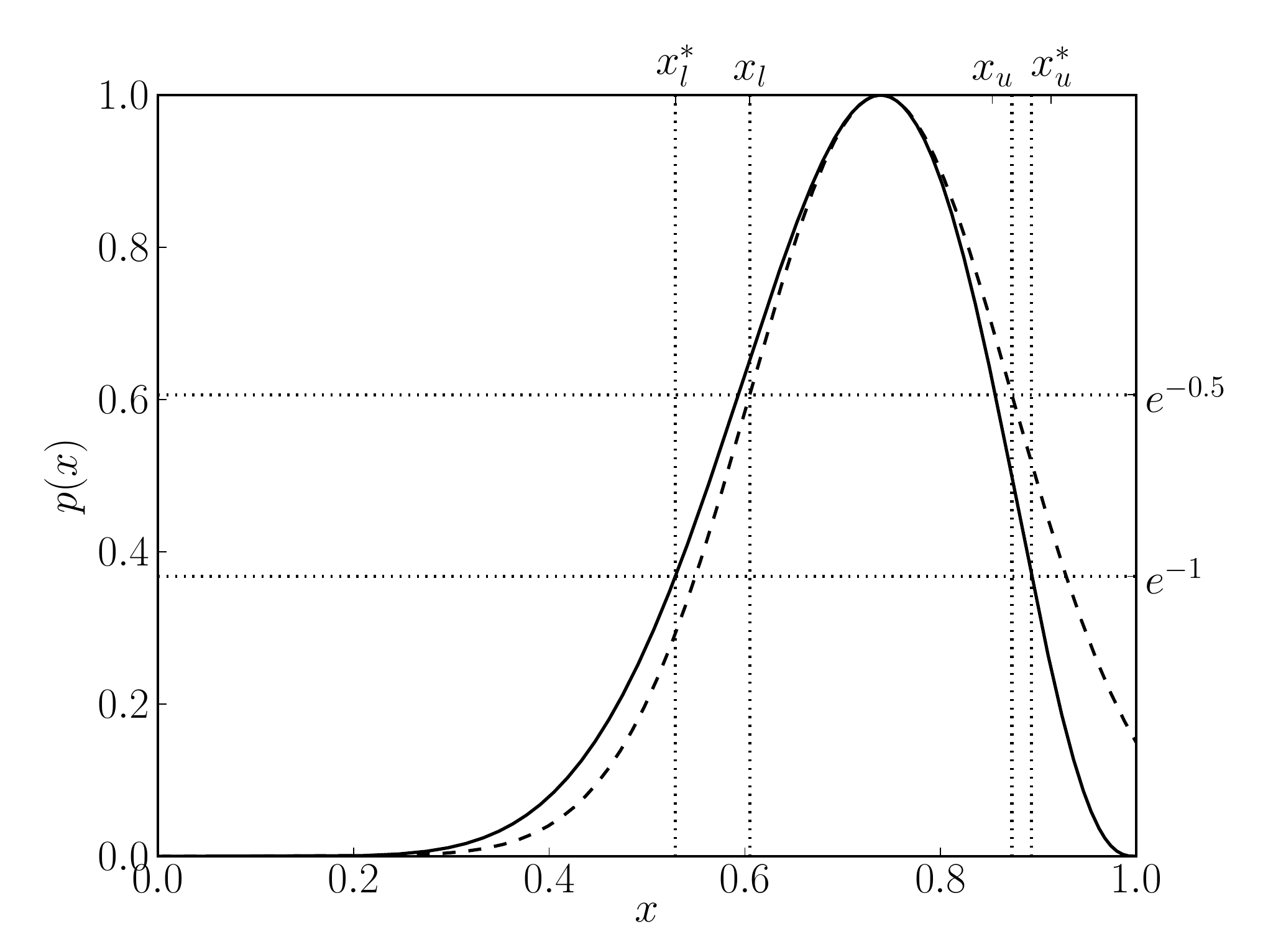}
  \caption{The picture shows the location of the optimal points $x_l^{*}$ and $x_u^{*}$ for $p(x)$ (solid line) and the points $x_l$ and $x_u$ obtained from the Gaussian approximation (dashed line). The Gaussian approximation touches the density at the mode point.}
\label{fig:optimal_points}
\end{figure}

Unfortunately we do not have the inverse to the conditional density \eqref{eqn:conditional_density}. We will use additional information about $p(x|a,b,c,d)$ to make a good although suboptimal choice for $x_l$ and $x_u$. In the following let $0<x_m<1$. A second order Taylor expansion of $q(x|a,b,c,d)$ around the mode point yields a Gaussian approximation of the conditional density, \[p(x)\approx\exp\left\{q(x_m)+\frac{q''(x_m)}{2}(x-x_m)^2\right\}.\] The standard deviation is given by  $\sigma=\sqrt{\frac{1}{-q''(x_m)}}$.  We simply set \begin{align*} &x_l=x_m-\sigma, & x_u=x_m+\sigma.\end{align*} A comparison between the optimal points and the ones obtained from the Gaussian approximation to $p(x)$ is shown in \autoref{fig:optimal_points}. 

\subsection{Rejection sampling using the envelope}
It is straightforward to generate random variates from the individual pieces of the enveloping function. There are fast and reliable implementations for the generation of uniform as well as for exponential random variates and we can use rejection to sample from pieces of $p(x)$ individually. We can decompose $p(x)$ as follows \begin{align*}p(x)= & p_1 p(x)\chi_{[0,x_l]}(x)+ p_2 p(x)\chi_{[x_l,x_u]}(x) + \\  &  p_3 p(x) \chi_{[x_u,1]}(x)\end{align*} with $p_i$ denoting the weights of the individual pieces. We do not know the discrete probabilities $p_i$ a priori but there is a simple and efficient algorithm circumventing the need for $p_i$ altogether. The algorithm does only require the weights $w_i$ of the individual pieces of $g(x)$. Due to the simple form of $g(x)$ obtaining analytic expressions for $w_i$ is straightforward. The following \autoref{algo:modified_composition} is a variant of the modified composition method that can be found in \cite[p. 69]{devroye1986}. In the following let $0<x_l<x_u<1$. The treatment of special cases is straightforward but requires additional branches in the algorithm complicating notation. The discrete probabilities of the individual dominating pieces $w_i$ are given by \begin{equation}\label{eqn:w_i}w_i=\frac{r_i}{\sum_{k=1}^{3}r_i},\end{equation} with \begin{equation}\label{eqn:r_1} r_1=\int_{-\infty}^{x_l} \mathrm{d}x \, p(x_l)e^{q'(x_l)(x-x_l)} = \frac{p(x_l)}{q'(x_l)}, \end{equation} \begin{equation}\label{eqn:r_2}r_2=\int_{x_l}^{x_u} \mathrm{d}x \, p(x_m)= p(x_m)(x_l-x_u),\end{equation} \begin{equation}\label{eqn:r_3}r_3=\int_{x_u}^{\infty} \mathrm{d}x \, p(x_l)e^{q'(x_l)(x-x_l)}=\frac{p(x_u)}{-q'(x_u)}.\end{equation} In order to increase numerical stability for cases in which $a$, $b$, $c$, and $d$ have large values \eqref{eqn:conditional_density} is usually scaled to the mode point value. Additionally the logarithm of the final acceptance condition in \autoref{algo:modified_composition}, \[\log(U)+h(x) \leq q(x),\] can be tested instead of the condition $U g(x) < p(x)$. The case $d \gg 1$ can for example occur in situations in which the probabilities for states differ by orders of magnitude, $\pi_i \ll \pi_j$, since $d \propto \pi_j/\pi_i$.
\begin{algorithm}
\DontPrintSemicolon
\SetAlgoNoLine
\SetInd{0.5em}{1.0em}
\KwIn{$a$, $b$, $c$, $d$}
\KwOut{$x$}
Compute $x_m$ using \eqref{eqn:mode_point}\;
$\sigma=\sqrt{\frac{1}{-q''(x_m)}}$\;
$x_l=x_m-\sigma$\;
$x_u=x_m+\sigma$\;
Compute $w_1$, $w_2$, and $w_3$ using \eqref{eqn:w_i}, \eqref{eqn:r_1}, \eqref{eqn:r_2}, \eqref{eqn:r_3}\;
\Repeat{$U g(x)<p(x)$}{
  $Z \sim \chi_{[0,1]}(x)$\;
  $U \sim \chi_{[0,1]}(x)$\;
  \uIf{$Z<w_1$}{
     \Repeat{$x \geq0$}{
        $y \sim e^{-q'(x_l)y}$\;
        $x=x_l-y$
     }
  }
  \uElseIf{$Z<w_1+w_2$}{
     $y \sim \chi_{[0,1]}(x)$\;
     $x=x_l+(x_u-x_l)y$
  }
  \Else{
     \Repeat{$x \leq 1$}{
        $y \sim e^{q'(x_u)y}$\;
        $x=x_l+y$\;
     }
  }
}
\caption{Sample $p(x|a,b,c,d)$ \label{algo:modified_composition}}
\end{algorithm}

\subsection{Modified rejection sampling for large $d$ values}
In the case $d \gg 1$ we can use an alternative strategy to generate samples according to \eqref{eqn:conditional_density}. We can rewrite 
\[p(x|a,b,c,d)\propto g(x|a,b) \psi(x|c,d)\]
with 
\[g(x|a,b)= \frac{\Gamma(a+b)}{\Gamma(a) \Gamma(b)} x^{a}(1-x)^{b}\]
and
\[\psi(x|c,d)=\left(\frac{d-x}{d}\right)^c.\]
The density $g(x|a,b)$ is the usual beta density which can be efficiently sampled and $\psi(x|c,d)$ is a $[0,1]$ valued function. The modified rejection method \cite{devroye1986}, \autoref{algo:modified_rejection}, can be used to generate samples from $p(x|a,b,c,d)$. The algorithm is efficient for cases in which $\psi(x|c,d) \approx 1$ for all $x \in [0,1]$. In the case $d \gg 1$ we obtain \[\psi(x|c,d)\approx 1-\frac{cx}{d}\] using a Taylor expansion in $x/d$. Since $x \in [0,1]$ the algorithm is efficient for  $c/d \ll 1$. It is straightforward to see that the efficiency of \autoref{algo:modified_rejection} increases with growing $d$. 
\begin{algorithm}
 \DontPrintSemicolon
\SetAlgoNoLine
\SetInd{0.5em}{1.0em}
\KwIn{$a$, $b$, $c$, $d$}
\KwOut{$x$}
\Repeat{$U<\psi(x|c,d)$}{
 $x \sim g(x|a,b)$ \;
 $U \sim \chi_{[0,1]}(x)$ \;
}
\caption{Modified rejection algorithm \label{algo:modified_rejection}}
\end{algorithm}

\section{Maximum likelihood estimation \label{sec:maximum_likelihood_estimation}}
The maximum likelihood estimate $T^*$ is the optimal point of the likelihood function $p(C|P)$. In other words the given observation $C$ is most likely to be generated by the optimal model $T^*$. The multinomial form of the likelihood and the linear nature of the constraints makes it possible to reformulate the problem of finding the maximum likelihood estimate as a convex optimization problem. Thus the global optimum $T^*$ can be efficiently found. For a thorough introduction and exhaustive overview see \cite{boyd2004}.

We note that $\log$ is a strictly monotone function so that finding the maximal point of $p(C|P)$ is equivalent to finding the maximal point of the log-likelihood function, \[l(C|P)=\log p(C|P).\] Finding the reversible transition matrix with given stationary distribution $\pi$ maximizing $l(C|P)$ can be stated as the following optimization problem.\begin{align*}
 & \text{minimize} & -\sum_{i,j=1}^{n} c_{ij} \log p_{ij}, & & \\
 & \text{subject to} & -p_{ij} \leq 0 & & 1\leq i,j \leq n, \\
 &                   & \sum_{k=1}^{n} p_{ik}=1 & & 1\leq i \leq n, \\
 &                   &  \pi_i p_{ij}-\pi_j p_{ji}=0 & & 1\leq i < j \leq n. 
\end{align*}
This is a constrained optimization problem in $n^2$ variables. A reduction in the number of independent variables can be achieved by eliminating constraints and explicitly incorporating them into the objective function and the remaining constraints. There exist a number of numerical libraries for the solution of convex optimization problems. We have used the freely available python cvxopt module \cite{anderson2011}. The numerical solution of a convex optimization problem is usually facilitated by iteratively updating the suboptimal point by solving a system of linear equations containing the first and second order derivatives of the objective function and all non-linear constraints as well as the matrices specifying the linear constraints. In order to start the iterative scheme one needs a valid initial point to start the iteration. In the following we will outline how we can compute a reversible transition matrix $P$ with fixed stationary distribution $\pi$ from any given possibly non-reversible transition matrix $Q$. Our method is similar to an approach outlined in \cite{jiang2009}. The guiding idea is the mechanism underlying the Metropolis-Hastings algorithm transforming an arbitrary transition matrix into one that is reversible with respect to a given stationary distribution. Denote by $a_{ij}$ the following weights, \begin{equation}\label{eqn:metropolis_weights} a_{ij}=\min\{1,\frac{\pi_j q_{ji}}{\pi_i q_{ij}}\}.\end{equation} These are precisely the weights in the Metropolis-Hastings algorithm. Define a new transition matrix $P^{(0)}$ by \begin{equation} \label{eqn:enforcing_detailed_balance} p_{ij}^{(0)}=\begin{cases} a_{ij}q_{ij} & i \neq j \\ 1-\sum_{k\neq i}a_{ik}q_{ik} & i=j \end{cases}.\end{equation} Observe that the diagonal elements $p_{ii}$ will always be greater after such a transformation $p_{ii}\geq q_{ii}$ for all $i=1,\dotsc,n$. We will use the transformation outlined above in order to generate a valid starting point for the likelihood maximization scheme. Since $Q$ is arbitrary we choose it to be the non-reversible maximum likelihood estimator, \[q_{ij}=\frac{c_{ij}}{\sum_{k=1}^{n}c_{ik}}.\] We generate$P^{(0)}$ by enforcing the reversibility condition with respect to $\pi$ using \eqref{eqn:metropolis_weights},\eqref{eqn:enforcing_detailed_balance}.

As an alternative, the likelihood maximization can be performed by iteratively maximizing the conditional probabilities of $p_{ij}$ for $i<j$. This is either done by the reversible transition matrix estimator described in \cite{prinz2011}, or by the following iterative algorithm, \autoref{algo:iterative_mle_fixed_pi}. The former algorithm is implemented in EMMA \cite{senne2012} by the \texttt{mm\_estimateFixedPi} command.
\begin{algorithm}
\DontPrintSemicolon
\SetAlgoNoLine
\SetInd{0.5em}{1.0em}
\KwIn{$\pi$, $C$}
\KwOut{$P^{*}$}
$P=P^{(0)}(\pi, C)$\;
$S=P$\;
\While{$\delta>\epsilon$}{
  \For{$i \in \{1,\dotsc,n\}$}{
    \For{$j \in \{1,\dotsc,n\}$}{
      \If{$i<j$}{
	Compute parameters $\Delta_{ij}$, $\Lambda_{ij}$, $a$, $b$, $c$, $d$.\;
	Mode point $x_m=x_m(a,b,c,d)$\;
        $p_{ij}=x_m \cdot \min(\Delta_{ij},\Lambda_{ij})$\;
        $p_{ii}=\Delta_{ij}-p_{ij}$\;
        $p_{ji}=\frac{\pi_i}{\pi_j}p_{ij}$\;
        $p_{jj}=\frac{\pi_i}{\pi_j}\Lambda_{ij}-p_{ji}$\;	
      }   
    }  
  }
  $\delta=|l(C|P) - l(C|S)|$\;
   $S=P$\;
}
$P^{*}=P$\;
\caption{Iterative maximum likelihood estimation with fixed stationary distribution \label{algo:iterative_mle_fixed_pi}}
\end{algorithm}

\section{A Gibbs sampler for transition matrices with fixed stationary distribution}
The Gibbs sampling approach as first presented in \cite{geman1984} achieves the following. Let $p(x_1,\dotsc,x_n)$ be a given joint probability distribution and denote by $p_i(x_i)$ the marginal distribution of the i-th variable, $p_i(x_i)=p(x_i|x_{j\neq i})$. It can be shown that under certain conditions (Lemma 10.11 in \cite{robert1999}) the ability to generate random variates from all conditionals $p_i(x_i)$ is sufficient to generate samples from the joint distribution $p(x_1,\dotsc,x_n)$. The algorithm can be stated as follows. Denote by $(x_1^{(k)},\dotsc,x_n^{(k)})$ the k-th random vector generated by the algorithm. Then a new sample is generated by ``sweeping'' through the vector updating all coordinates from the respective conditional densities. In other words, for $i$ in $1,\dotsc,n$,
\[x_i^{(k+1)} \sim p(x_i|x_1^{(k+1)},\dotsc,x_{i-1}^{(k+1)},x_{i+1}^{(k)},\dotsc,x_n^{(k)}).\]
There exist several variants of the Gibbs sampling algorithm, the ``random scan'' version which picks $i$ from $\{1,\dotsc,n\}$ at random and returns a new sample after $n$ of such updates instead of sweeping through all coordinates in succession is especially popular. 

Recall that the conditional distribution of $p_{ij}$ is given by \eqref{eqn:conditional_density_raw} with parameters $\Delta_{ij}$ and $\Lambda_{ij}$ explicitly given by \eqref{eqn:def_delta} and \eqref{eqn:def_gamma}. Having obtained an explicit expression for $p(p_{ij}|p_{k\neq i,j,l\neq i,j})$ for all $i<j$  we can proceed to construct a Gibbs sampling algorithm to generate random variates $P$ from the joint distribution. In order to start the Markov chain we need a valid initial transition matrix $P^{(0)}$ obeying detailed balance with respect to the given stationary distribution $(\pi_i)_{1\leq i\leq n}$. The matrix $P^{(0)}$ should also be irreducible so that choosing any irreducible transition matrix $q_{ij}$ and enforcing detailed balance with respect to $\pi$ according to \eqref{eqn:enforcing_detailed_balance} will result in a valid initial transition matrix possessing the desired properties. However, we recommend to use the maximum likelihood estimate obtained above as a starting point for the Gibbs chain to immediately draw transition matrices from regions of high probabilities. The computation of parameters during the Gibbs sampling procedure can be simplified by noting that \[\Delta_{ij}^{(k)}=p_{ij}^{(k-1)}+p_{ii}^{(k-1)},\] \[\Lambda_{ij}^{(k)}=\frac{\pi_j}{\pi_i}\left(p_{jj}^{(k-1)}+p_{ji}^{(k-1)}\right),\] where $k$ denotes the step in the Gibbs sampling chain. In other words coupling between elements is mediated only by diagonal elements $p_{ii}$. This gives rise to \autoref{algo:Gibbs_sampling_fixed_stationary_distribution}. The usual procedure returns only after $l$ such elementary steps have been taken resulting in a sequence $P^{(0)},P^{(l)},\dotsc,P^{(Nl)}$ of transition matrices. Here $l$ denotes the number of independent variables, $l=n(n-1)/2$.
\begin{algorithm}
\DontPrintSemicolon
\SetAlgoNoLine
\SetInd{0.5em}{1.0em}
\KwIn{$\pi$, $C$, $P^{(k-1)}$}
\KwOut{$P^{(k)}$}
\Repeat{$i<j$}
{
$i \sim \{1,\dots,n\}$ \;
$j \sim \{1,\dots,n\}$ \;
}
$\Delta_{ij}^{(k)}=p_{ij}^{(k-1)}+p_{ii}^{(k-1)}$\;
$\Lambda_{ij}^{(k)}=\frac{\pi_j}{\pi_i}\left(p_{jj}^{(k-1)}+p_{ji}^{(k-1)}\right)$\; 
\eIf{
  $\Delta_{ij}^{(k-1)} \leq \Lambda_{ij}^{(k-1)}$}{$d=\frac{\Delta_{ij}^{(k-1)}}{\Lambda_{ij}^{(k-1)}}$\;
  $b=C_{ii}$\;
  $c=C_{jj}$\;
}
{
  $d=\frac{\Lambda_{ij}^{(k-1)}}{\Delta_{ij}^{(k-1)}}$\;
  $b=C_{jj}$ \;
  $c=C_{ii}$ \;
}
Sample $x \sim x^{a}(1-x)^b(d-x)^c$ using \autoref{algo:modified_composition}\;
$p_{ij}^{(k)}=x \cdot \min(\Delta_{ij}^{(k-1)},\Lambda_{ij}^{(k-1)})$ \;
$p_{ii}^{(k)}=\Delta_{ij}^{(k-1)}-p_{ij}^{(k)}$ \;
$p_{ji}^{(k)}=\frac{\pi_i}{\pi_j}p_{ij}^{(k)}$ \;
$p_{jj}^{(k)}=\frac{\pi_i}{\pi_j} \Lambda_{ij}^{(k-1)}-p_{ji}^{(k)}$\;
\caption{Gibbs sampling of $P$ with fixed stationary distribution $\pi$ \label{algo:Gibbs_sampling_fixed_stationary_distribution}}   
\end{algorithm}

In the case that $\pi$ has some uncertainty specified by the probability model \eqref{eqn:probability_stationary_distribution}, the generation of a compatible ensemble of transition matrices can be achieved using \eqref{eqn:probability_stationary_and_transitions} given that samples of $\pi$ can be generated according to $p(\pi|E)$. The following \autoref{algo:Gibbs_sampling_uncertain_stationary_distribution} generates $P\sim p(P|C,E)$. The number $k$ is usually taken as the minimal number of runs to decorrelate from the starting point $P^{(0)}$. 
\begin{algorithm}
\DontPrintSemicolon
\SetAlgoNoLine
\SetInd{0.5em}{1.0em}
\KwIn{$C$, $E$, $k$}
\KwOut{$P$}
$\pi \sim p(\pi|E)$\;
Compute $P^{(0)}$ according to \eqref{eqn:enforcing_detailed_balance} or via \autoref{algo:iterative_mle_fixed_pi}\;
\For{$i \in \{1,\dotsc,k\}$}{
Generate $P^{(i)}$ via \autoref{algo:Gibbs_sampling_fixed_stationary_distribution} using $\pi$, $C$, $P^{(i-1)}$\;
}
$P=P^{(k)}$\;
 \caption{Gibbs sampling of $P$ with uncertain stationary distribution \label{algo:Gibbs_sampling_uncertain_stationary_distribution}}
\end{algorithm}

\subsection{Enforcing sparsity - a prior for metastable dynamics}
The equilibrium dynamics of proteins does often exhibit the feature of metastability. Thus any transition matrix characterizing an approximation via a Markov jump process on conformation space should also exhibit traits of metastability. As discussed in \cite{deuflhard2000} metastable Markov processes on discrete state spaces can be understood in terms of nearly uncoupled Markov chains with small transition probabilities between blocks defining the dynamics within a single metastable subset. For finite observations of the metastable process the small probabilities for transitions between metastable sets and the zero probabilities of forbidden transitions might become indistinguishable in an ensemble of transition matrix generated by a sampling approach with no prior information. If the uncertainties of small but non-zero transition probabilities are of the same order than those that correspond to forbidden transitions it might not be possible to recover the desired metastable properties from the generated ensemble. In physical systems there are of course no forbidden transitions since all transition probabilities are strictly positive. However many of these might be still orders of magnitude smaller than the transition probabilities between meta stable regions. In practice we will not observe any of such transitions in a finite realization of our process, not even for a typical realization long enough to achieve sufficient sampling of meta stable transitions. In this case we can treat them, in a very good approximation, as forbidden transitions.

We will show how one can enforce the generation of an ensemble of transition matrices compatible with the sparsity structure of the given observations. If one assumes detailed balance with respect to $\pi$ observed transitions $c_{ij}$ as well as observed transitions $c_{ji}$ indicate nonzero probabilities $p_{ij}$, $i<j$. Therefore we conclude that whenever $c_{ij}+c_{ji}>0$ the probability of $p_{ij}>0$ is positive, $\mathbb{P}(p_{ij}>0)>0$. To enforce sampling of metastable transition matrices we require that $p_{ij}=0$ if $c_{ij}+c_{ji}=0$, $i<j$, for all $P$ in the sample. In other words the sparsity structure of $C+C^{T}$ is enforced for all $P$. This sparsity prior is equivalent to using a prior count of -1 on all $(i,j)$ with $c_{ij}+c_{ji} = 0$ in \eqref{eqn:probability_transition_matrix} (see supplementary information of \cite{noe2009}). The sparse prior is applied by restricting the sampling algorithm to those elements $p_{ij}$ for which $c_{ij}+c_{ji}>0$. It is apparent that the \[q_{ij}=\frac{c_{ij}+c_{ji}}{\sum_{k}(c_{ik}+c_{ki})}\] possesses the desired sparse structure. Furthermore, generating $P^{(0)}$ according to \eqref{eqn:enforcing_detailed_balance} does not change the sparsity structure of the off-diagonal elements of $Q$. Starting from a transition matrix with the desired sparse structure it is straightforward to restrict the Gibbs sampling algorithm by updating only elements for which $i<j$ and $c_{ij}+c_{ji}>0$. Denote by 
\[\theta=\{(i,j) \in \mathbb{N}^2|1\leq i<j \leq n,\, c_{ij}+c_{ji}>0\}\] In \autoref{algo:Gibbs_sampling_fixed_stationary_distribution_sparse} we outline a method to generate a sample of reversible transition matrices with fixed stationary distribution obeying the sparse structure given by $\theta$.
\begin{algorithm}
\DontPrintSemicolon
\SetAlgoNoLine
\SetInd{0.5em}{1.0em}
\KwIn{$\pi$, $C$, $P^{(k-1)}$, $\theta$}
\KwOut{$P^{(k)}$}
Draw $(i,j)$ uniformly from $\theta$ \;
Proceed as in \autoref{algo:Gibbs_sampling_fixed_stationary_distribution} \;
\caption{Gibbs sampling of $P$ with fixed stationary distribution $\pi$ - Sparse version \label{algo:Gibbs_sampling_fixed_stationary_distribution_sparse}}   
\end{algorithm}

\section{Results}
In the following we will show that the above outlined Gibbs sampling converges much faster than the Metropolis Hastings approach developed in \cite{noe2008}. Please recall that the general Metropolis Hastings approach generates random variates from the density $p(x)$ by choosing proposals $y$ conditioned on the current state of the chain $x$ from a proposal density $q(x,y)$ and accepts proposed samples with the following acceptance probability, \[a(x,y)=\min\{1,\frac{p(y)q(y,x)}{p(x)q(x,y)}\}.\] The crucial difference between Gibbs sampling and Metropolis Hastings sampling is that the Metropolis chain remains in the current state as long as the proposed value is rejected while the Gibbs sampling approach generates a new sample at each step. This possibility to remain in the current state usually leads to longer correlation times for the Metropolis chain than for the Gibbs chain. Thus one needs to run longer Metropolis chains than Gibbs chains to achieve an equal degree of convergence. On the other hand one needs to be able to generate random variates from all conditionals efficiently while the Metropolis chain can be advanced using a possibly very simple proposal density $q(x,y)$. In the following we will compare our current sampling approach with the one developed in \cite{noe2008} and demonstrate improved convergence and more-rapidly decaying autocorrelation. We start with a simple model using the following count matrix \begin{equation} \label{eqn:count_matrix_model_system} C=\left(\begin{array}{lll} 100 & 5 & 0 \\ 20 & 4 & 20 \\ 0 & 8 & 75 \end{array} \right)\end{equation} and stationary distribution \begin{equation}\label{eqn:stationary_distribution_model_system} \pi=\left(0.5,0.01,0.49\right).\end{equation} to assess the convergence properties of the two approaches.

\subsection{Conditional distributions}
\begin{figure*}[htb]
      \subfloat[]{
        \includegraphics[width=0.3\textwidth]{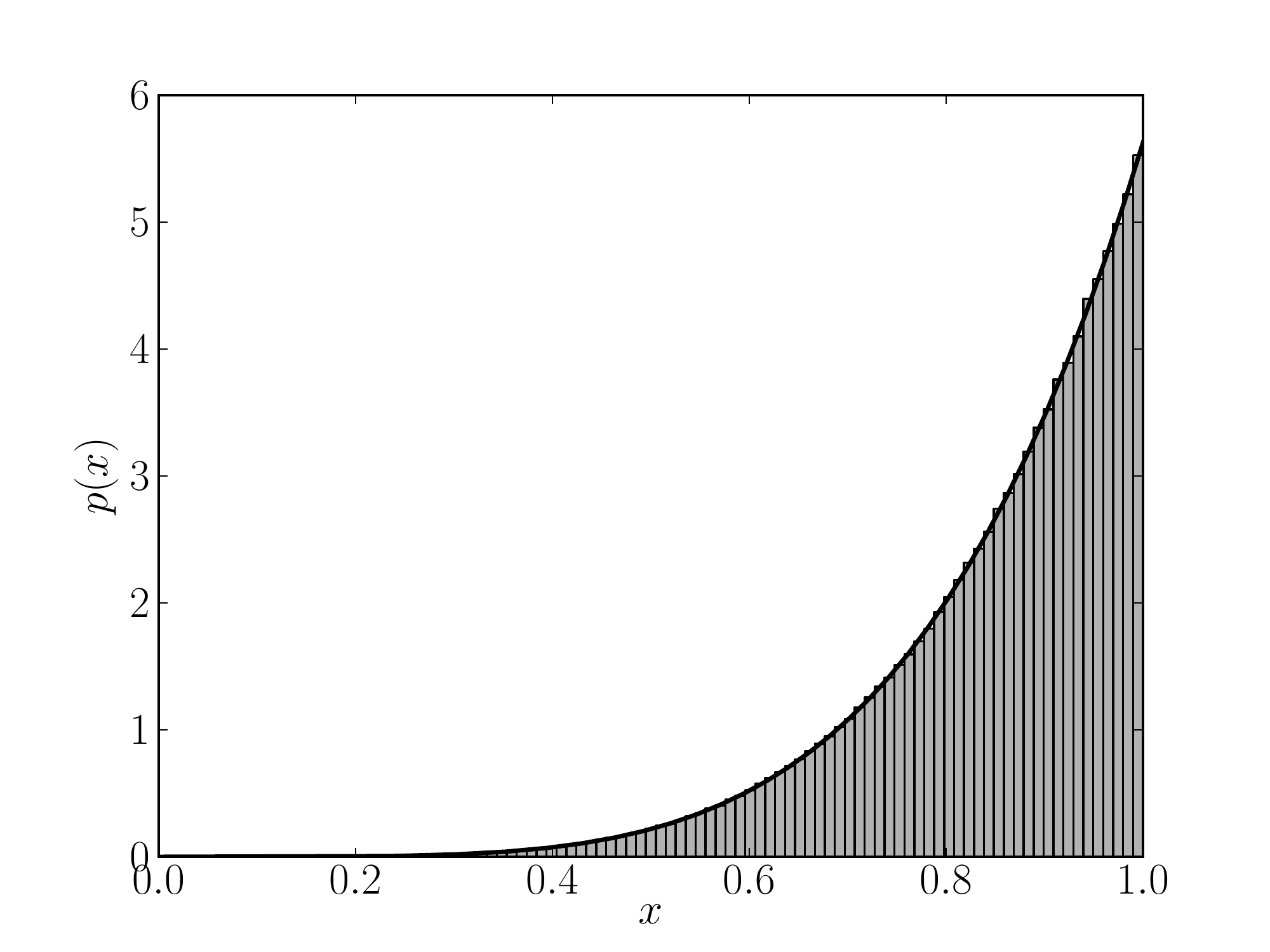}
        \label{fig:a5_b0_c4_d10}
      }
      \subfloat[]{
         \includegraphics[width=0.3\textwidth]{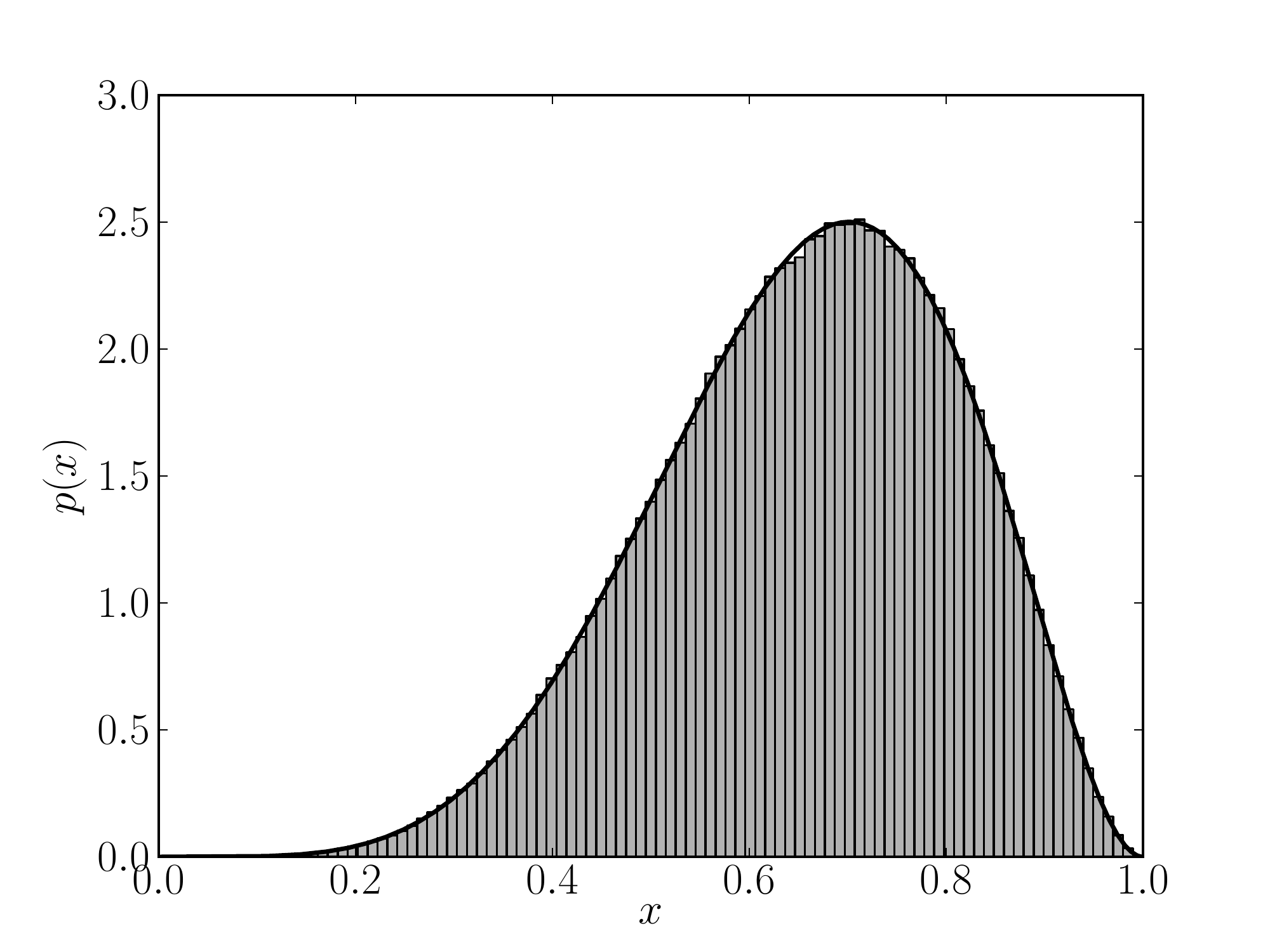}
         \label{fig:a5_b2_c4_d10}
      }
      \subfloat[]{
         \includegraphics[width=0.3\textwidth]{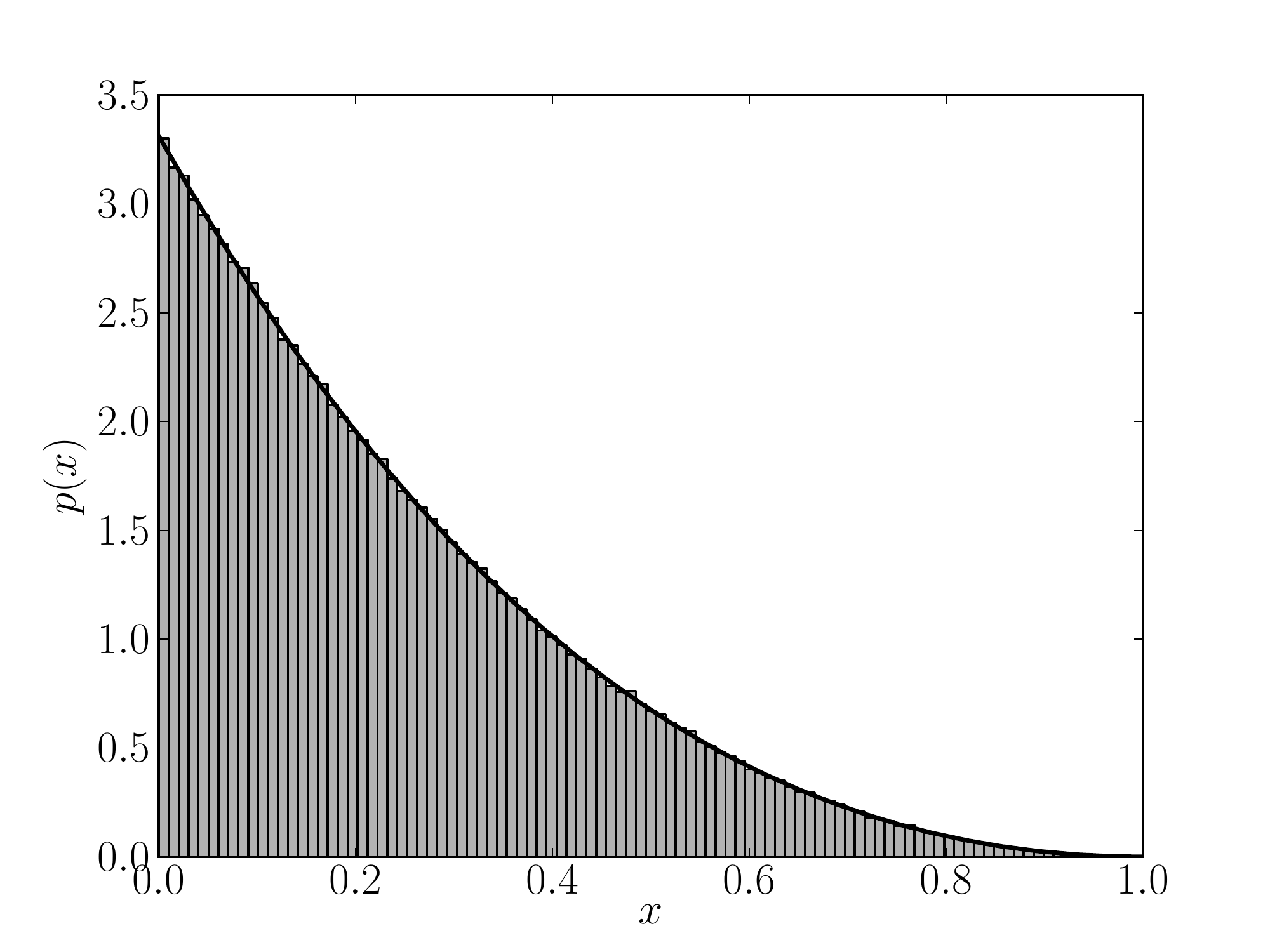}
         \label{fig:a0_b2_c4_d10}
      }

        \subfloat[]{
        \includegraphics[width=0.3\textwidth]{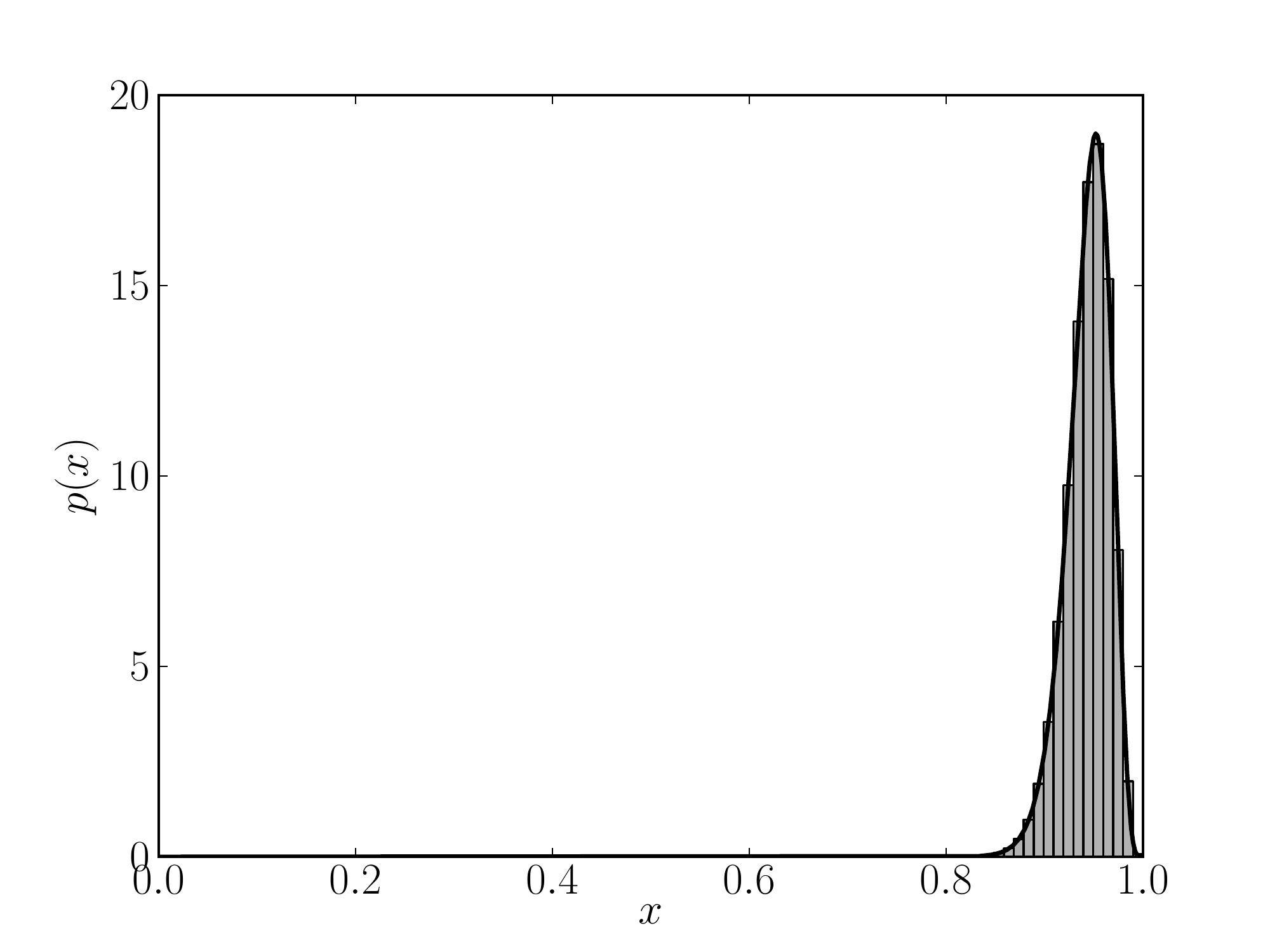}
        \label{fig:a100_b5_c40_d100}
      }
      \subfloat[]{
         \includegraphics[width=0.3\textwidth]{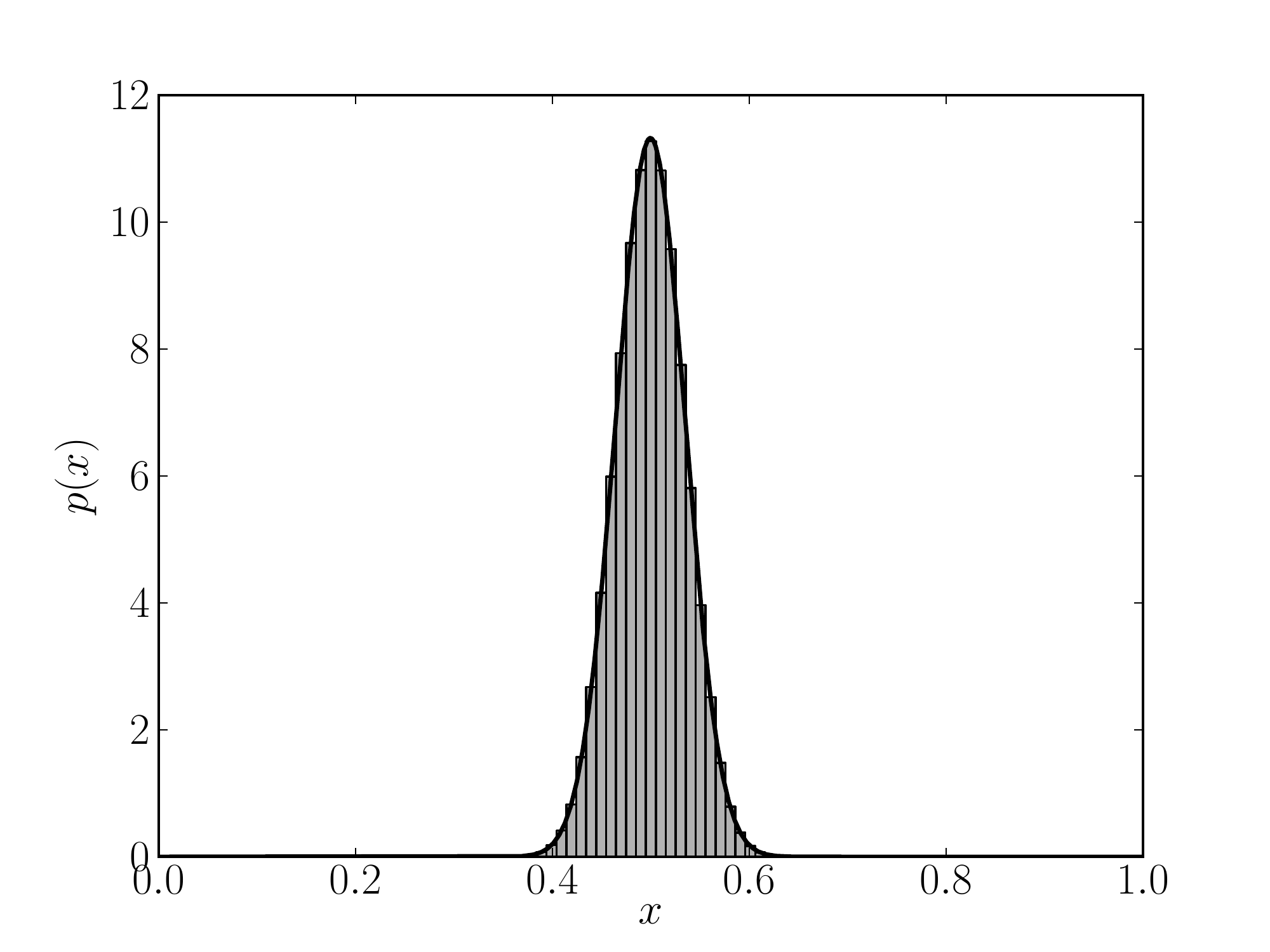}
         \label{fig:a100_b100_c40_d100}
      }
      \subfloat[]{
         \includegraphics[width=0.3\textwidth]{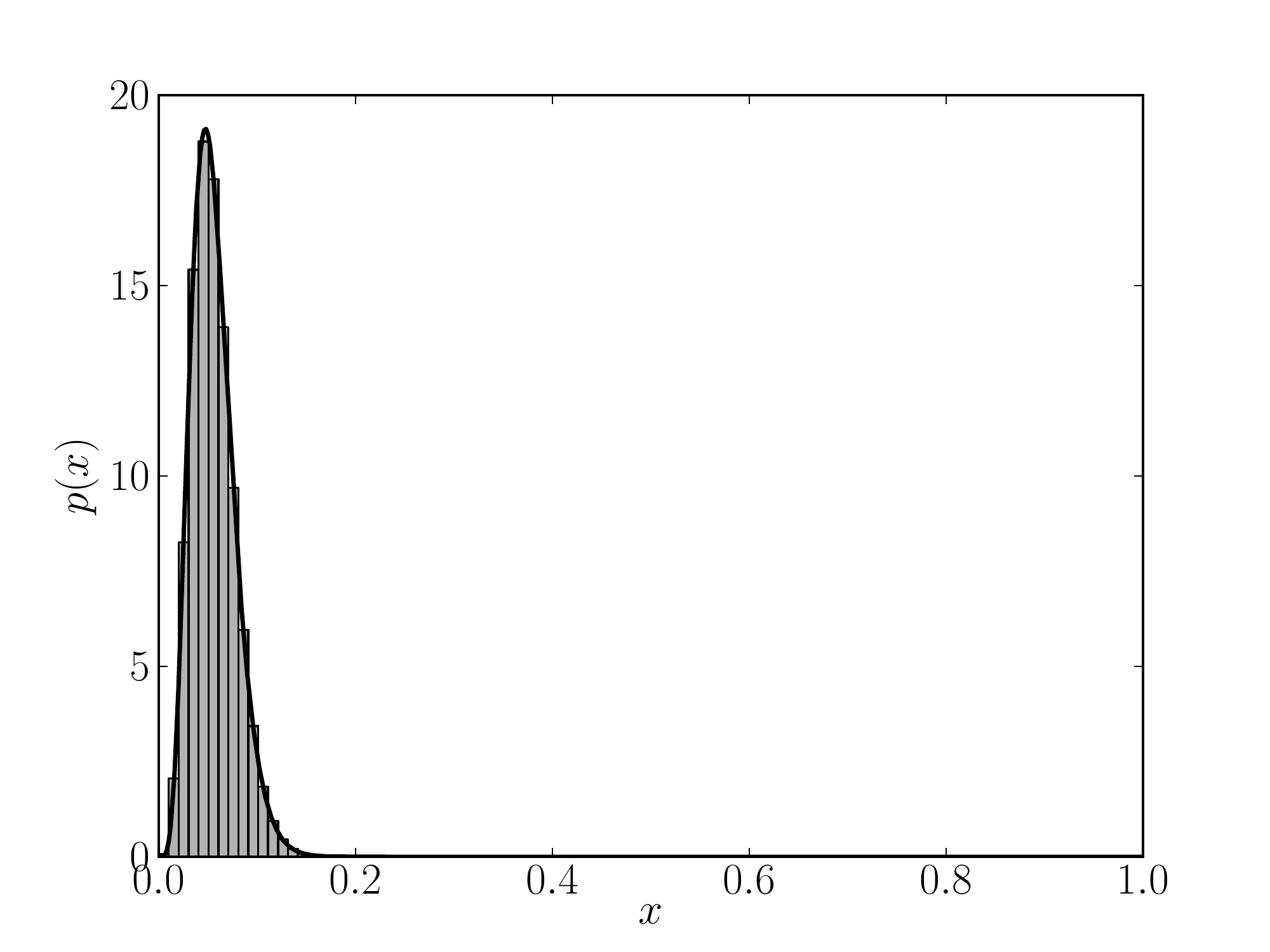}
         \label{fig:a5_b100_c40_d100}
      }     
      
      \subfloat[]{
        \includegraphics[width=0.3\textwidth]{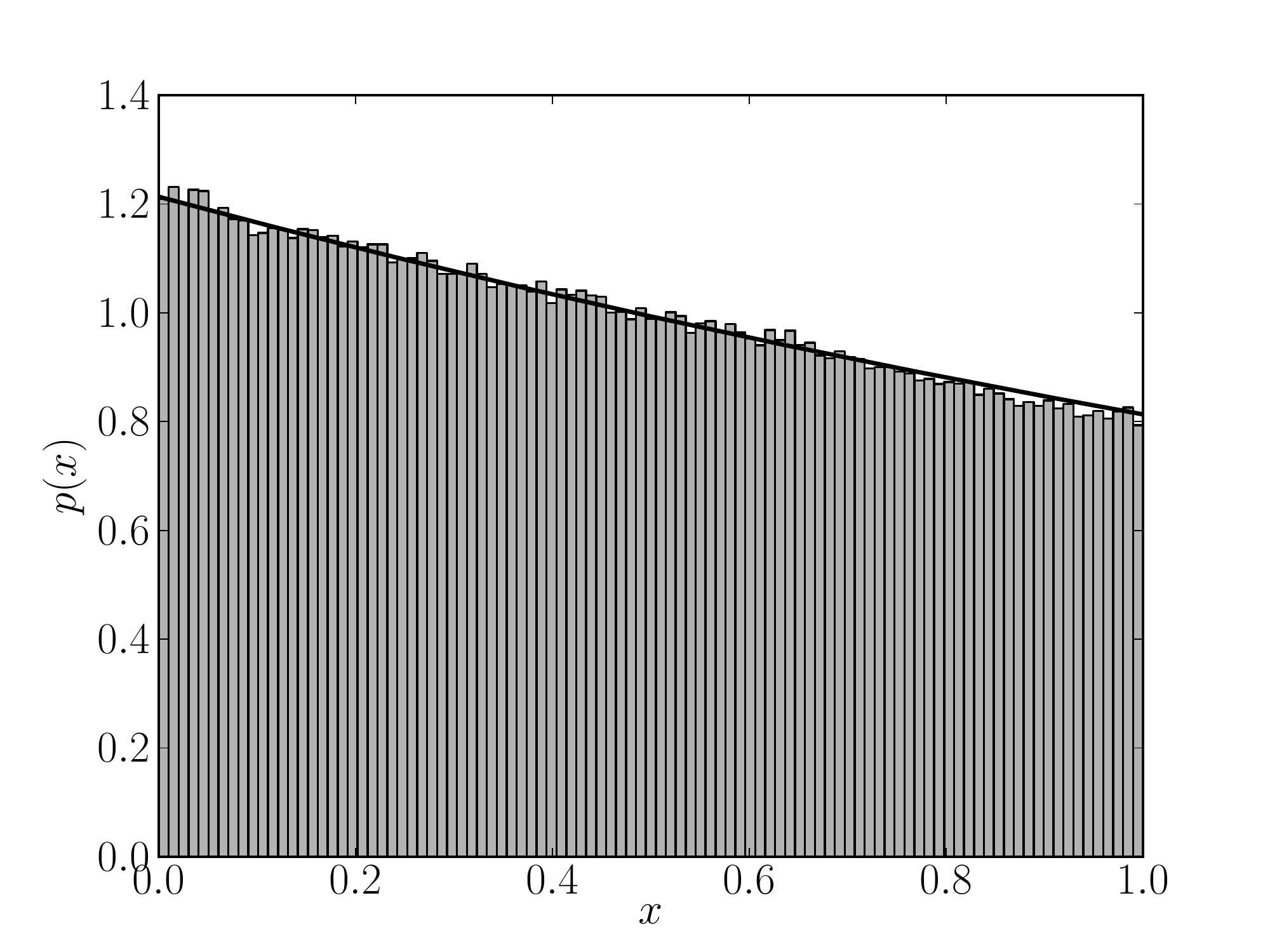}
        \label{fig:a0_b0_c4_d10}
      }
      \subfloat[]{
         \includegraphics[width=0.3\textwidth]{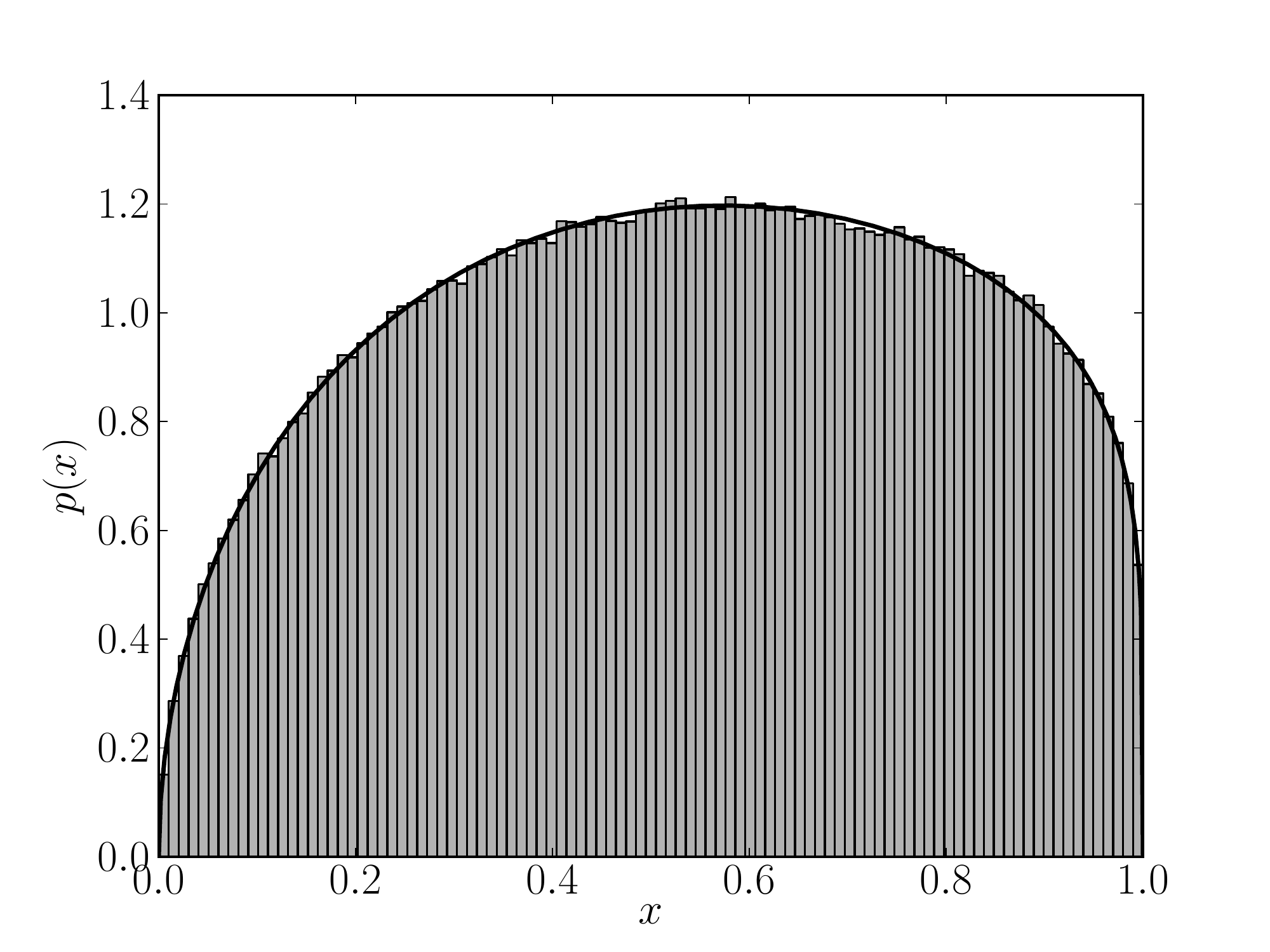}
         \label{fig:a05_b02_c40_d100}
      }
      \subfloat[]{
         \includegraphics[width=0.3\textwidth]{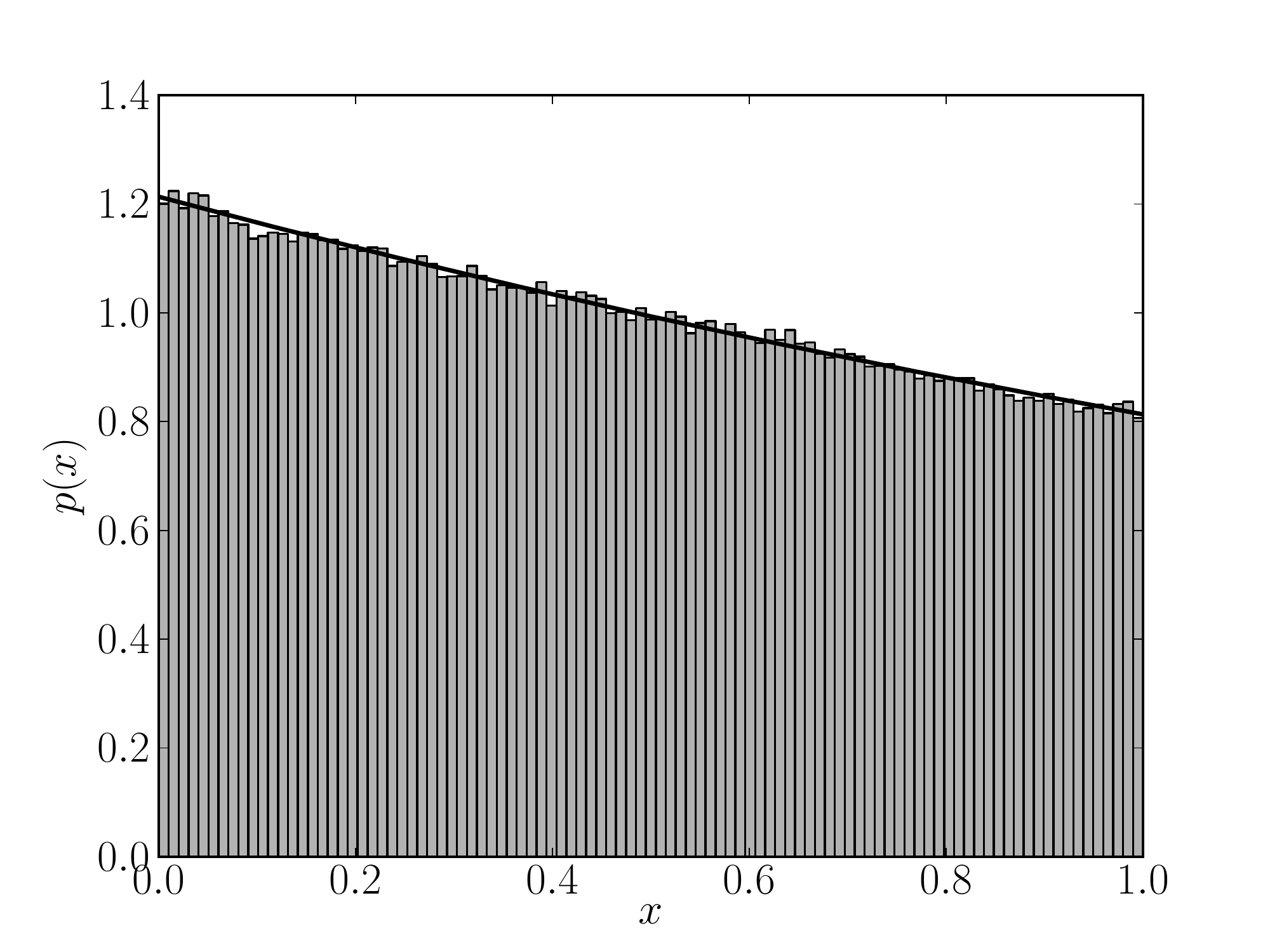}
         \label{fig:a0_b0_c4k_d10k}
      }   
\caption{Conditional density $p(x)$ (solid line) for different parameters $a$, $b$, $c$, and $d$. The histograms show a sample of $N=10^6$ random variates generated using the method outlined above. First row $c=4$, $d=10$: (a) $a=5$, $b=0$, (b) $a=5$, $b=2$, (c) $a=0$, $b=2$. Second row $c=40$, $d=100$: (d) $a=100$, $b=5$, (e) $a=100$, $b=100$, (f) $a=5$, $b=100$. Third row: (g) $a=0$, $b=0$, $c=4$, $d=10$, (h) $a=0.5$, $b=0.2$, $c=40$, $d=100$, (i) $a=0$, $b=3 \cdot 10^4$, $c=4\cdot 10^{3}$, $d=10^4$.}
\label{fig:shape_density}
\end{figure*}

We have generated a sample of $N=10^6$ random variates from the conditional density, $p(x|a,b,c,d)$ \eqref{eqn:conditional_density}, for various choices of parameters $a$, $b$, $c$, $d$ to demonstrate the ability of our new method to correctly generate random variates from all possible conditional densities. In \autoref{fig:shape_density} we compare the shape of each histogram to the graph of the exact density function for the same parameter values. The figures clearly indicate that all densities have been correctly sampled.

\subsection{Convergence of mean values and variances}
\begin{figure*}
    \subfloat[]{
      \includegraphics[width=0.45\textwidth]{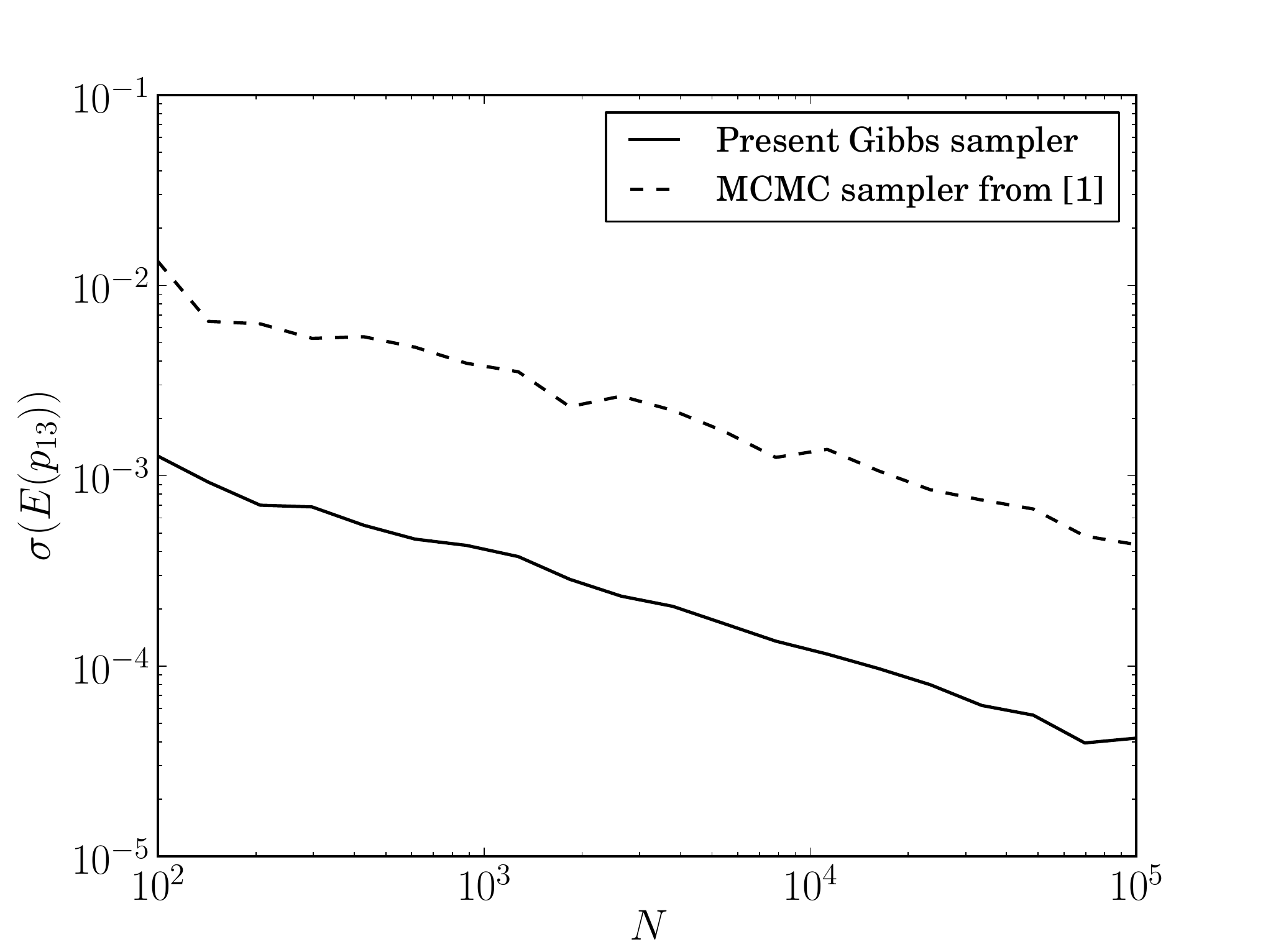}
      \label{fig:std_old_vs_new_Ep13_loglog}
    }
    \subfloat[]{
      \includegraphics[width=0.45\textwidth]{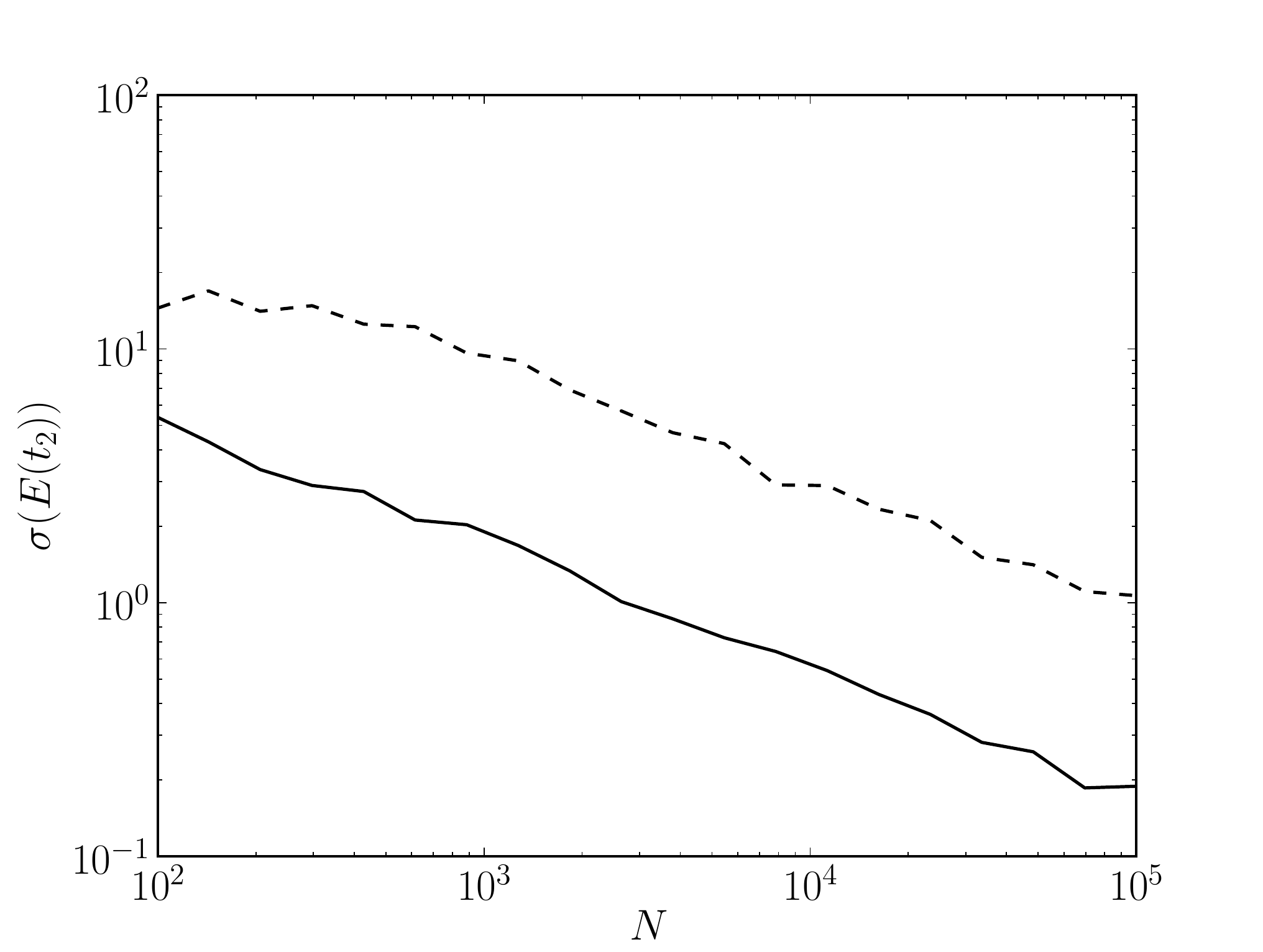}
      \label{fig:std_old_vs_new_Et2_loglog}
    }

    \subfloat[]{
      \includegraphics[width=0.45\textwidth]{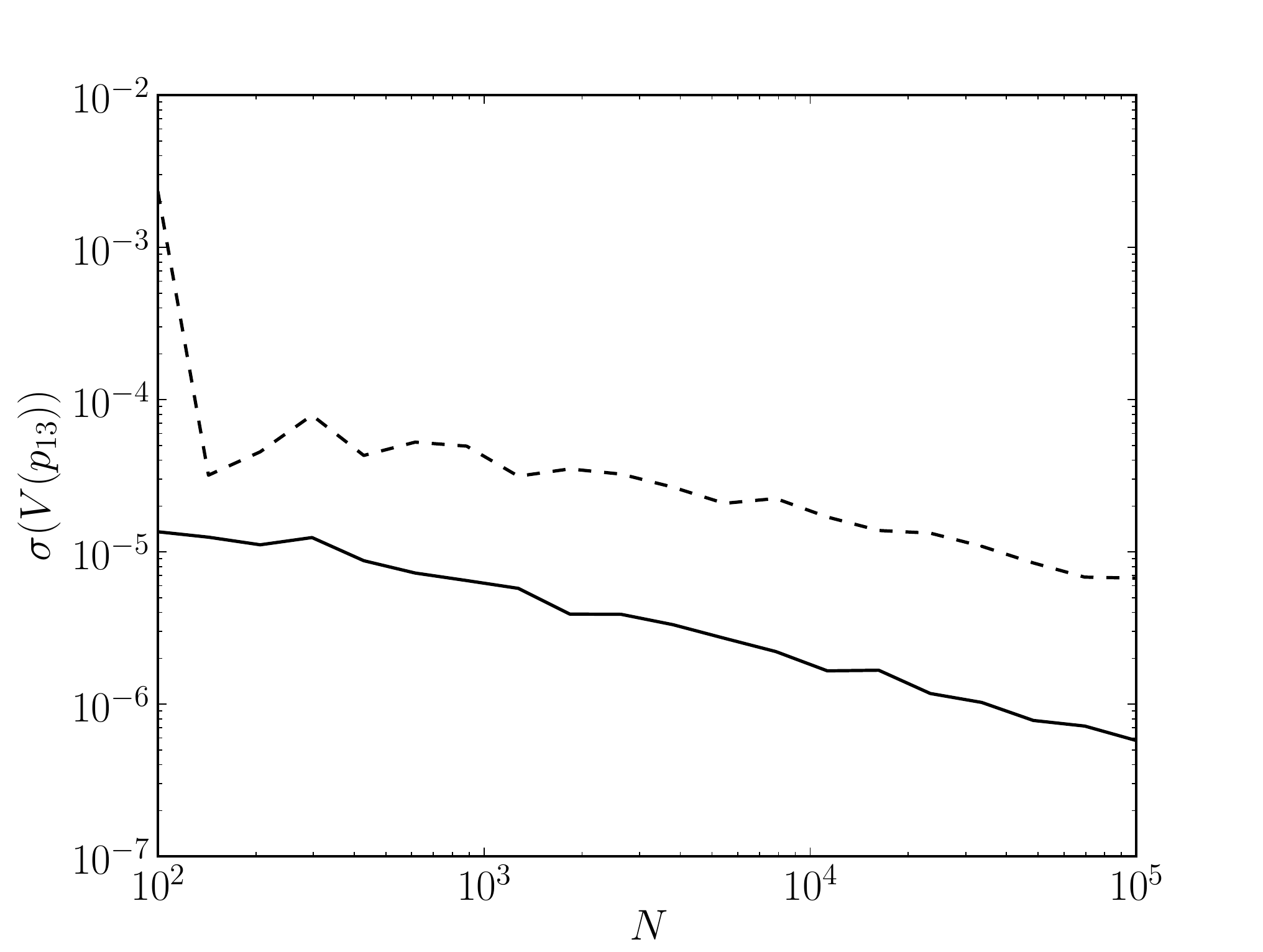}
      \label{fig:std_old_vs_new_Vp13_loglog}
    }
    \subfloat[]{
      \includegraphics[width=0.45\textwidth]{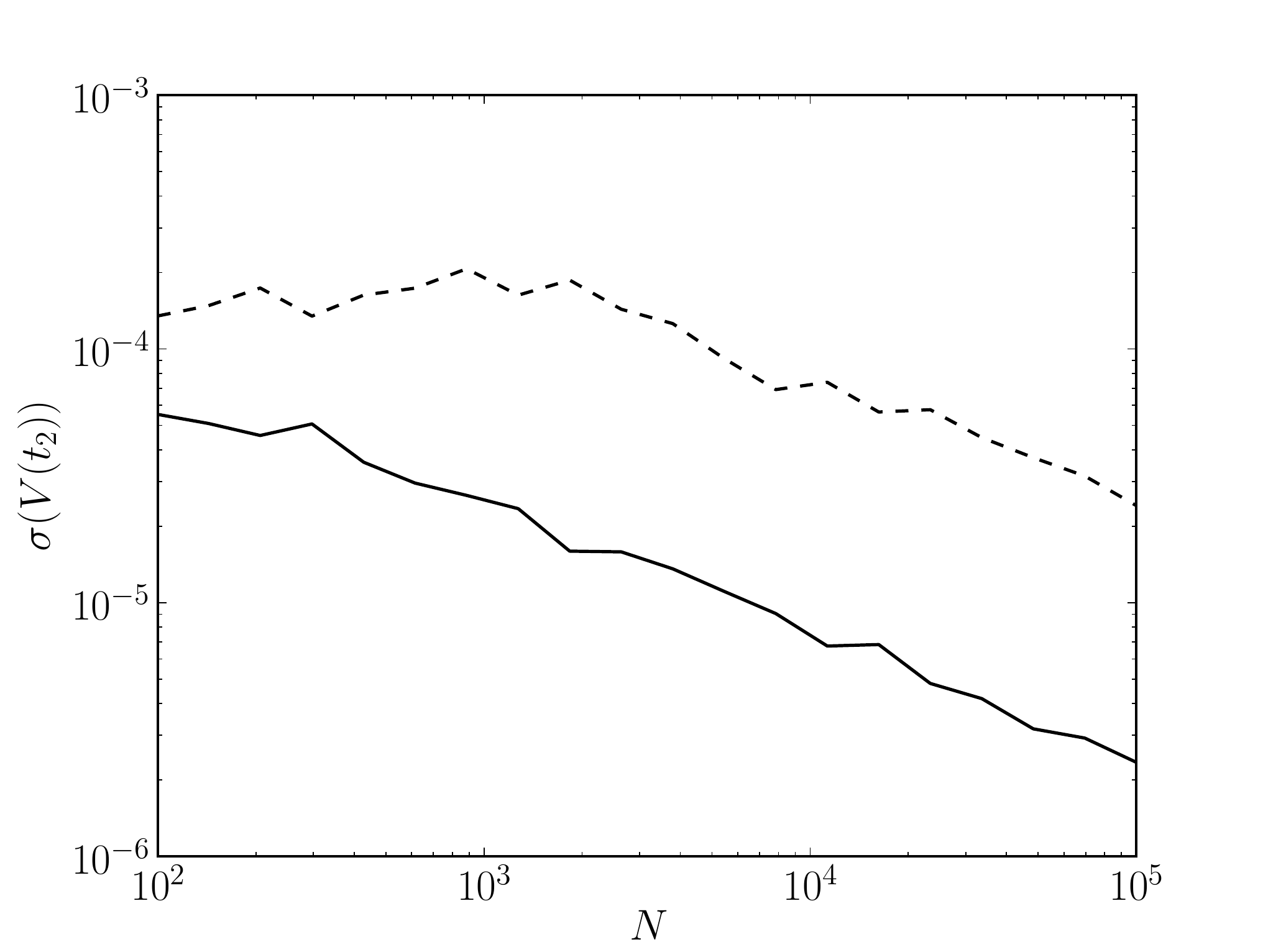}
      \label{fig:std_old_vs_new_Vt2_loglog}
    }
    \caption{Results obtained for the model system with count matrix \eqref{eqn:count_matrix_model_system} and stationary distribution \eqref{eqn:stationary_distribution_model_system}. Standard deviation for estimated mean and variance of observables is plotted against the number of elementary sampling steps $N$. (a) mean transition matrix element $\mathbb{E}(p_{13})$, (b) mean of the second largest implied time scale $\mathbb{E}(t_2)$, (c) transition matrix element variance $\mathbb{V}(p_{13})$, (d) variance of the second largest implied time scale $\mathbb{V}(t_2)$. The Gibbs sampler introduced here (solid line) convergences faster than the Metropolis chain from \cite{noe2008} (dashed line) by almost two orders of magnitude. For the mean second largest time scale $\mathbb{E}(t_2)$, (b), the achieved speedup is more than one order of magnitude.}
    \label{fig:old_vs_new_mean_and_var}
\end{figure*}

In order to assess the quality of a Monte Carlo sampling procedure one usually computes the standard error of the mean of an observable $O$ estimated from a finite sample generated by evolving the chain for a finite number of steps. As observable we choose the value of individual matrix elements, $O=p_{ij}$ and the value of the second largest implied time scale, $O=t_2$. We have generated a maximum likelihood reversible transition matrix of the count matrix \eqref{eqn:count_matrix_model_system} with stationary distribution \eqref{eqn:stationary_distribution_model_system} using the algorithm in \cite{senne2012}. Then $n_{\text{ensemble}}=100$ independent Gibbs samplers using \autoref{algo:Gibbs_sampling_fixed_stationary_distribution} were used, taking $N$ steps in the range $10^2 \dots 10^5$, estimating $\mathbb{E}(p_{ij})$ and $\mathbb{E}(t_2)$ for each $(n_{\text{ensemble}},N)$. Then we have estimated the standard deviation over the sample for each (fixed) $N$. See Figure\autoref{fig:std_old_vs_new_Ep13_loglog} for a comparison of the convergence of $\mathbb{E}(p_{ij})$ between the two sampling approaches. The slowest relaxation timescale $t_2$, Figure\autoref{fig:std_old_vs_new_Et2_loglog}, is an example for a global observable with a functional dependence on all elements $p_{ij}$ so that the expectation value $\mathbb{E}(t_2)$ is a suitable measure to access the convergence of general observables. A comparison of the convergence behavior is shown in Figure\autoref{fig:std_old_vs_new_Et2_loglog}. We can also choose to observe the variance of individual matrix elements as well as the variance of the second largest implied timescale. The setup is identical to the one outlined for mean values. The figure also shows the convergence of the variance $\mathbb{V}(x)$ for a single matrix element, Figure\autoref{fig:std_old_vs_new_Vp13_loglog}, as well as for the implied timescale, Figure\autoref{fig:std_old_vs_new_Vt2_loglog}. The figures clearly indicate the improved convergence properties of the presented approach over the previous MCMC sampler in \cite{noe2008}, requiring two orders of magnitude less sampling steps to achieve a similar error level. 

\subsection{Autocorrelation functions}
\begin{figure*}
    \subfloat[]{
      \includegraphics[width=0.45\textwidth]{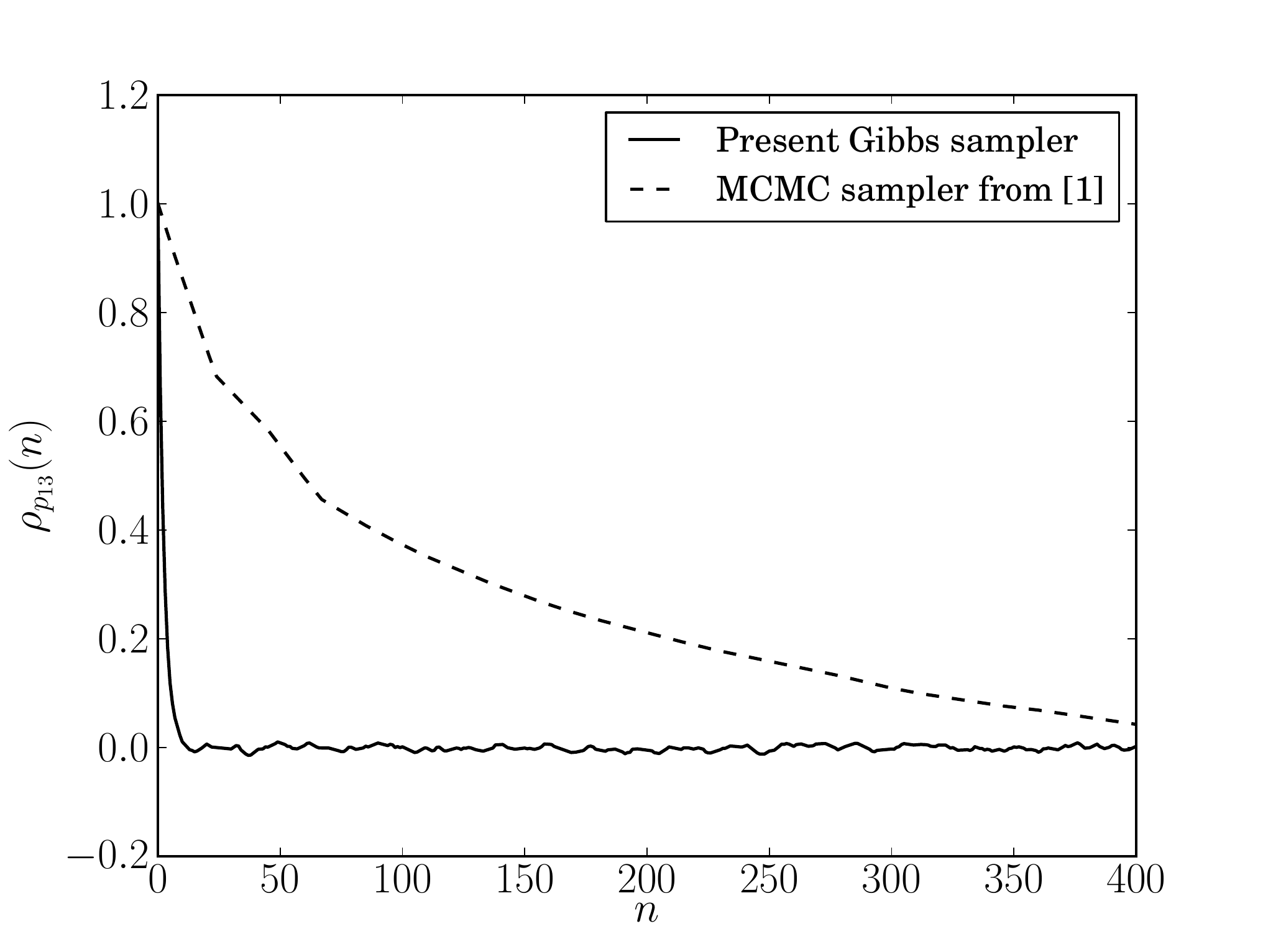}
      \label{fig:acf_old_vs_new_p13}
    }
    \subfloat[]{
      \includegraphics[width=0.45\textwidth]{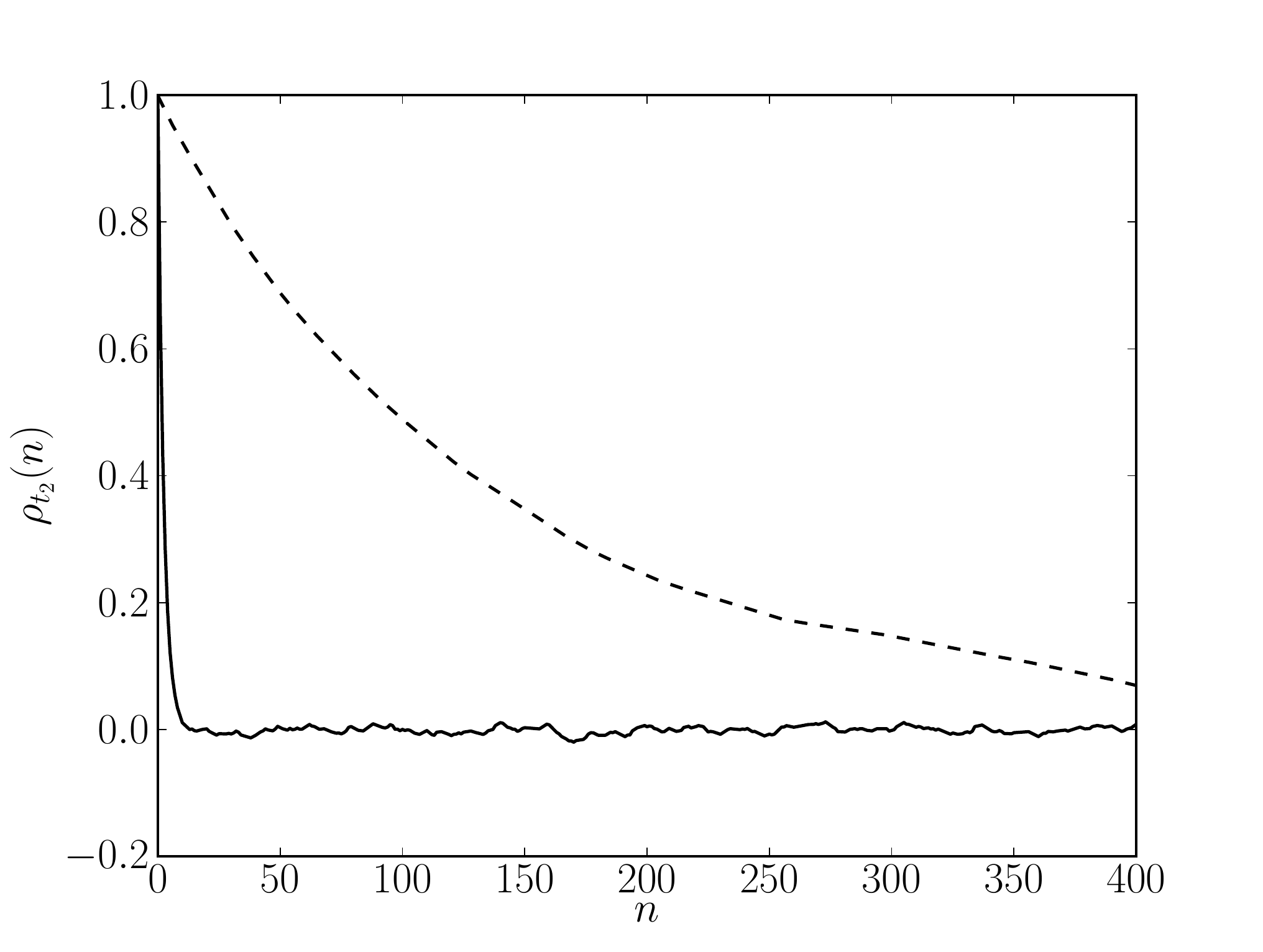}
      \label{fig:acf_old_vs_new_t2}
    }
\caption{Autocorrelation functions for the model system with count matrix \eqref{eqn:count_matrix_model_system} and stationary distribution \eqref{eqn:stationary_distribution_model_system}. (a) autocorrelation function for the transition matrix element $p_{13}$, (b) autocorrelation function for the second largest implied time scale $t_2$. The number of steps to take until samples are decorrelated $n_{decorr}$ is two orders of magnitude smaller for the Gibbs sampling method (solid line), $n_{decorr}=3$ for $p_{13}$ as well as for $t_{2}$, compared to the Metropolis sampling method (dashed line), $n_{decorr}=123$ for $p_{13}$ and $n_{decorr}=135$ for $t_2$.}
\label{fig:old_vs_new_autocorrelation}
\end{figure*}

As another measure of the improved convergence properties we compare the mixing times of the MCMC chain and the Gibbs chain. We have generated a sample of $10^5$ transition matrices for the above count matrix, \eqref{eqn:count_matrix_model_system}, and stationary distribution, \eqref{eqn:stationary_distribution_model_system}. Each sample was generated by advancing the chain using a single Gibbs or Metropolis step. Let $X_k$ be the value of the observable for the k-th sample. We have estimated the normalized autocorrelation function \[\rho_{X}(n)=\frac{\mathbb{E}[(X_k-\mu)(X_{k+n}-\mu)]}{\sigma^2}\] using the following estimator for the sample autocorrelation  \[\rho_{X}(n)=\frac{1}{\sigma^2(N-l)}\sum_{k=0}^{N-1-l} (X_k-\mu)(X_{k+n}-\mu),\] with $0\leq n\leq l$. The autocorrelation function for $p_{13}$ in Figure\autoref{fig:acf_old_vs_new_p13} clearly indicates the faster decay of autocorrelations for the Gibbs sampler. The autocorrelation function for the second largest implied time scale demonstrates a significant improvement over the previous approach, see \autoref{fig:acf_old_vs_new_t2}. The number of steps required in order to generate decorrelated samples, $n_{decorr}$, has been estimated by assuming an exponential decay for the autocorrelation function, \[\rho(n)=e^{-\frac{n}{n_{decorr}}}.\] The area under the graph of the autocorrelation function was computed using the trapezoidal rule and used as an estimate for $n_{decorr}$. Values for $n_{decorr}$ as well as for the corresponding decorrelation time, $t_{decorr}$, for observables $p_{13}$ and $t_2$ can be found in \autoref{tab:decorrelation_times}. $n_{decorr}$ is two orders of magnitude smaller for the Gibbs sampler than for the Metropolis approach. Due to the comparable speed of elementary sampling steps for both algorithms the improved decorrelation constant, $n_{decorr}$, leads to a similar improvement in decorrelation time.

\subsection{Application to simulation data}
\begin{figure*}
  \subfloat[]{
    \includegraphics[width=0.45\textwidth]{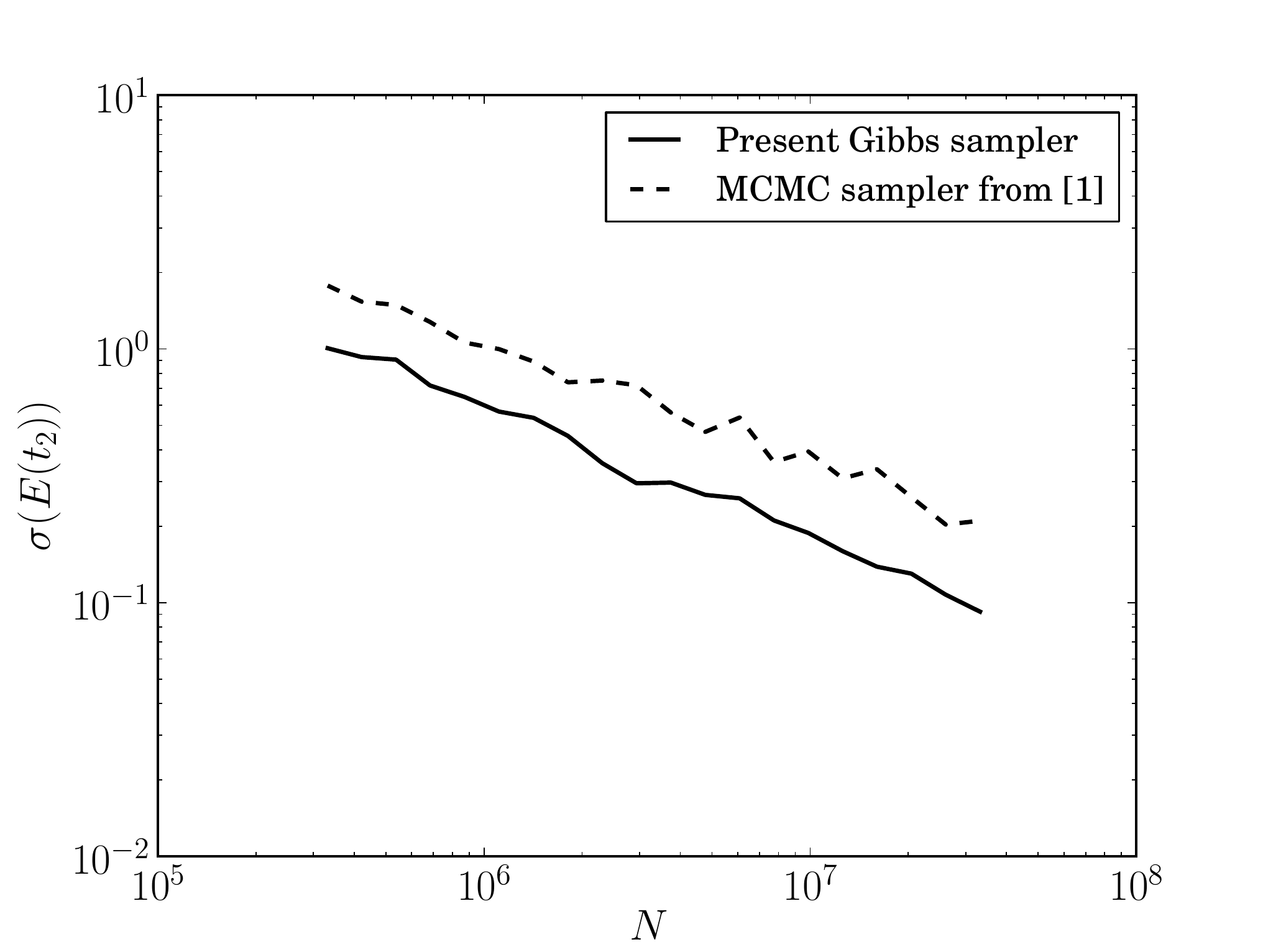}
      \label{fig:std_old_vs_new_Et2_GSGS}
  }
\subfloat[]{
    \includegraphics[width=0.45\textwidth]{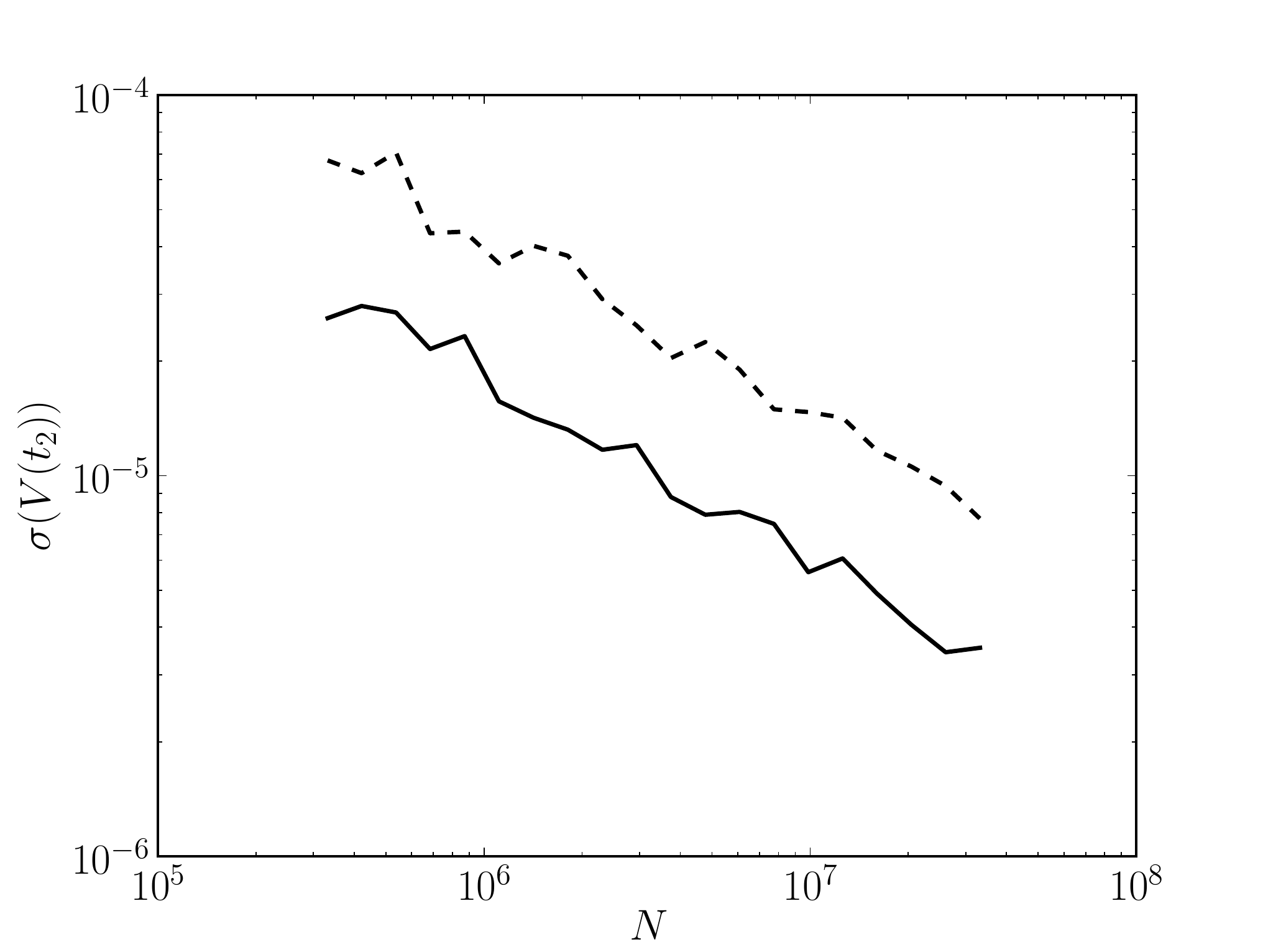}
      \label{fig:std_old_vs_new_Vt2_GSGS}
  }
\caption{Results obtained for the synthetic peptide MR121-GSGS-W.  Standard deviation of \autoref{fig:std_old_vs_new_Et2_GSGS} the mean implied time scale $\mathbb{E}(t_2)$, \autoref{fig:std_old_vs_new_Vt2_GSGS} the implied time scale variance $\mathbb{V}(t_2)$. The Gibbs sampler (solid line) shows a faster convergence than the Metropolis sampler (dashed line) for mean and variance of the second largest implied timescale $t_2$.}
\label{fig:old_vs_new_GSGS}
\end{figure*}

\begin{figure}
 \includegraphics[width=0.45\textwidth]{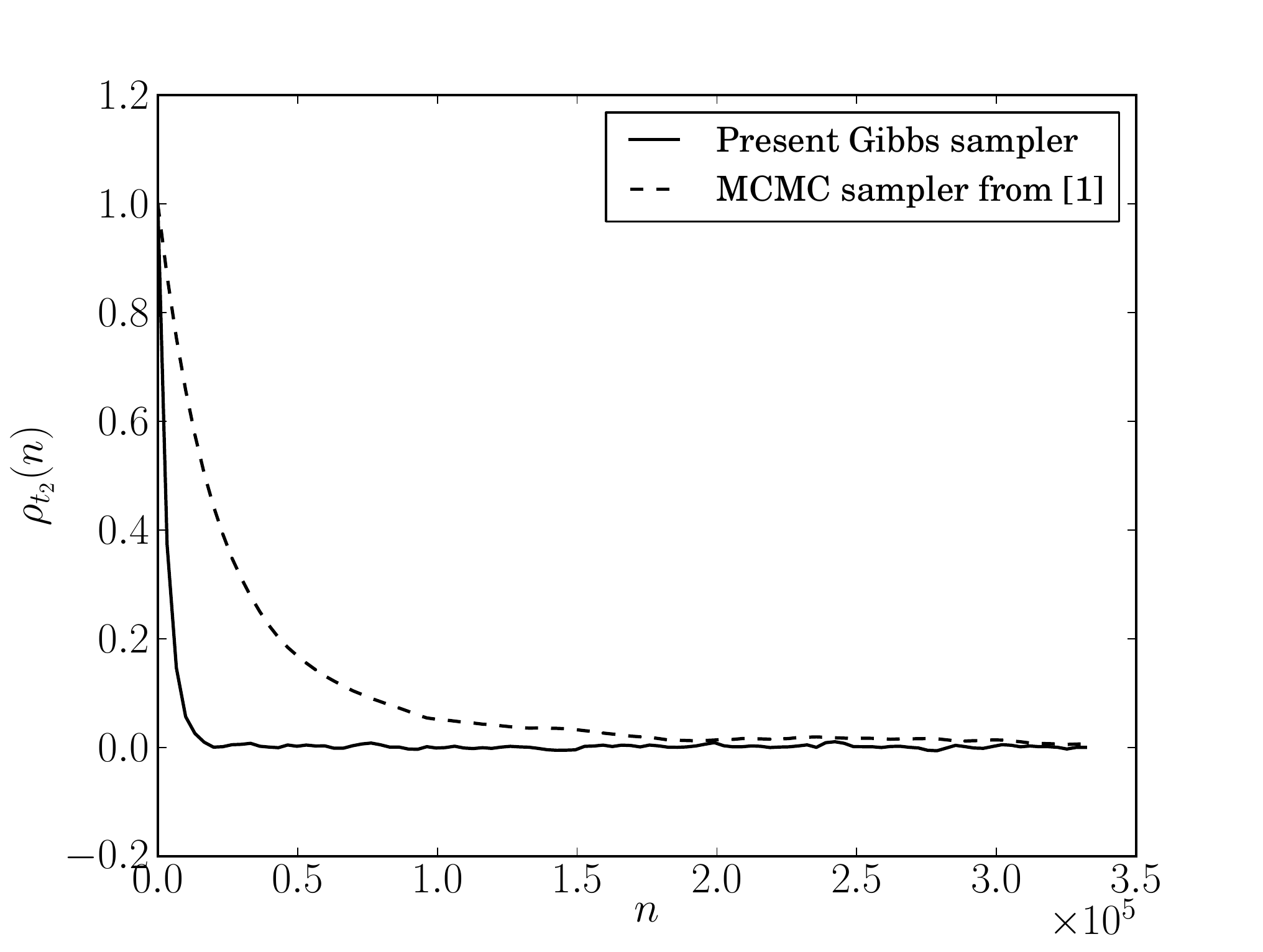}
\caption{Autocorrelation function for the MR121-GSGS-W peptide count matrix. The number of steps to take until samples are decorrelated $n_{decorr}$ is an order of magnitude smaller for the Gibbs sampling method (solid line), $n_{decorr}=4600$, compared to the Metropolis sampling method (dashed line), $n_{decorr}=33000$.}
 \label{fig:old_vs_new_acf_GSGS}
\end{figure}

In order to demonstrate the performance of the transition matrix sampling method we have applied the presented transition matrix Gibbs sampling algorithm to simulation data for the synthetic peptide MR121-GSGS-W. Trajectories were obtained by standard equilibrium dynamics simulations of a constant volume ensemble at $293K$ in explicit water with the Berendson thermostat using the Gromacs\cite{lindahl2001} simulation software. Each of the two trajectories used has a total length of $4 \mu s$ with trajectory frames separated by a time step of $10ps$. A detailed description of the simulation setup can be found in the supplementary information of \cite{noe2011}. The trajectories were clustered using regular spatial clustering of RMSD distances using EMMA \cite{senne2012}. A spatial cutoff of $3.5nm$ resulted in a clustering with $107$ distinct micro states. In order to obtain an estimate for the stationary probabilities of each micro state a Markov model with a lag time of $10ns$ was generated using the reversible transition matrix estimator presented in \cite{prinz2011}. The stationary distribution was obtained from the estimated transition matrix as the left eigenvector with eigenvalue $1$. A corresponding matrix containing transition counts between individual micro states was obtained by counting transitions at the same lag time. The sparse prior for metastable dynamics presented above was used to generate an ensemble of transition matrices using both, the Metropolis, and the Gibbs sampling procedure. We have started $n_{\text{ensemble}}=100$ independent chains with $N$ steps in the range $10^5 \dots 10^7$ and estimated $\mathbb{E}(t_2)$ and $\mathbb{V}(t_2)$ for each $(n_{\text{ensemble}},N)$. The standard deviation was estimated over the sample for each (fixed) $N$. In order to speed up the computation we have estimated $t_2$ from a spectral decomposition only after $l$ elementary sampling steps. We have chosen $l$ as the number of non zero independent transition probabilities $p_{ij}$. \autoref{fig:old_vs_new_GSGS} clearly indicates the improved convergence properties of the presented approach. Here, the Gibbs procedure needs one order of magnitude less sampling steps to reach a similar error level. A comparison of the autocorrelation functions for $t_2$, \autoref{fig:old_vs_new_acf_GSGS}, shows an order of magnitude smaller decorrelation constant $n_{decorr}$ for the Gibbs sampler compared to the Metropolis sampler . \autoref{tab:decorrelation_times} shows $n_{decorr}$ with the corresponding decorrelation time $t_{decorr}$. Due to the comparable speed of elementary sampling steps for both algorithms the improved decorrelation constant, $n_{decorr}$, leads to a one order of magnitude lower decorrelation time, $t_{decorr}$, for the Gibbs sampling algorithm.

\subsection{Computational efficiency}
For large entries in the count matrix $c_{ij}\gg 1$ the affected conditionals will be sharply peaked so that the uniform proposal densities used in \cite{noe2008} will have very low acceptance rates. This results in slow mixing chains and large autocorrelation times for the Metropolis algorithm so that much longer chains have to be run in order to achieve the same level of convergence. Due to the fact that \autoref{algo:Gibbs_sampling_fixed_stationary_distribution} only use standard distributions to envelope the conditionals, generating a random variate by rejection sampling from the conditional density is efficient enough to allow the generation of long chains. The Gibbs sampling algorithm, \autoref{algo:Gibbs_sampling_fixed_stationary_distribution}, was implemented using the colt library \cite{hoschek2000}. The algorithm performs $10^{8}$ elementary sampling steps in $89.1s$ on a 2GHz Intel processor. Performing the same number of elementary Metropolis steps takes $83.4s$, with a $27\%$ overall acceptance rate. The acceptance rate for the Metropolis step is also highly dependent on the specific element. The number of steps required in order to generate decorrelated samples $n_{decorr}$ as well as the wall-clock decorrelation time $t_{decorr}$ is shown in \autoref{tab:decorrelation_times} as an indicator of computational efficiency. The decorrelation time is calculated as $n_{decorr} \cdot t_{sample}$ with $t_{sample}=0.9 \mu s$ for the Gibbs sampling algorithm $t_{sample}=0.8 \mu s$ for the Metropolis algorithm.
\begin{table}
\begin{ruledtabular}
\begin{tabular}{rrrrr}

 & \multicolumn{2}{c}{Gibbs sampler} & \multicolumn{2}{c}{Metropolis sampler} \\
 & $n_{decorr}$ & $t_{decorr}$ & $n_{decorr}$ & $t_{decorr}$  \\[0.3cm] 
\multicolumn{5}{c}{$3\times3$ count matrix from \eqref{eqn:count_matrix_model_system}} \\[0.1cm]

$p_{13}$  & $3$  & $2.7 \mu s$ & $123$ & $98.4 \mu s$  \\ 
$t_2$ & $3$ &  $2.7 \mu s$ & $135$ & $108 \mu s$\\ [0.3cm]
\multicolumn{5}{c}{MR121-GSGS-W peptide} \\[0.1cm] 

$t_2$ & $4600$ & $4.14 ms$  &   $33000$ & $26.4ms$ \\

\end{tabular}
\end{ruledtabular}
\caption{Decorrelation times for estimated mean and variances of observables. The results were generated for the model system with count matrix \eqref{eqn:count_matrix_model_system} and stationary distribution \eqref{eqn:stationary_distribution_model_system} ($p_{13}$ and $t_2$) and for the MR121-GSGS-W peptide ($t_2$ only).}
\label{tab:decorrelation_times}
\end{table}

\section{Comparison with nonreversible and reversible estimation}
\begin{figure*}
  \subfloat[]{
    \includegraphics[width=0.33\textwidth]{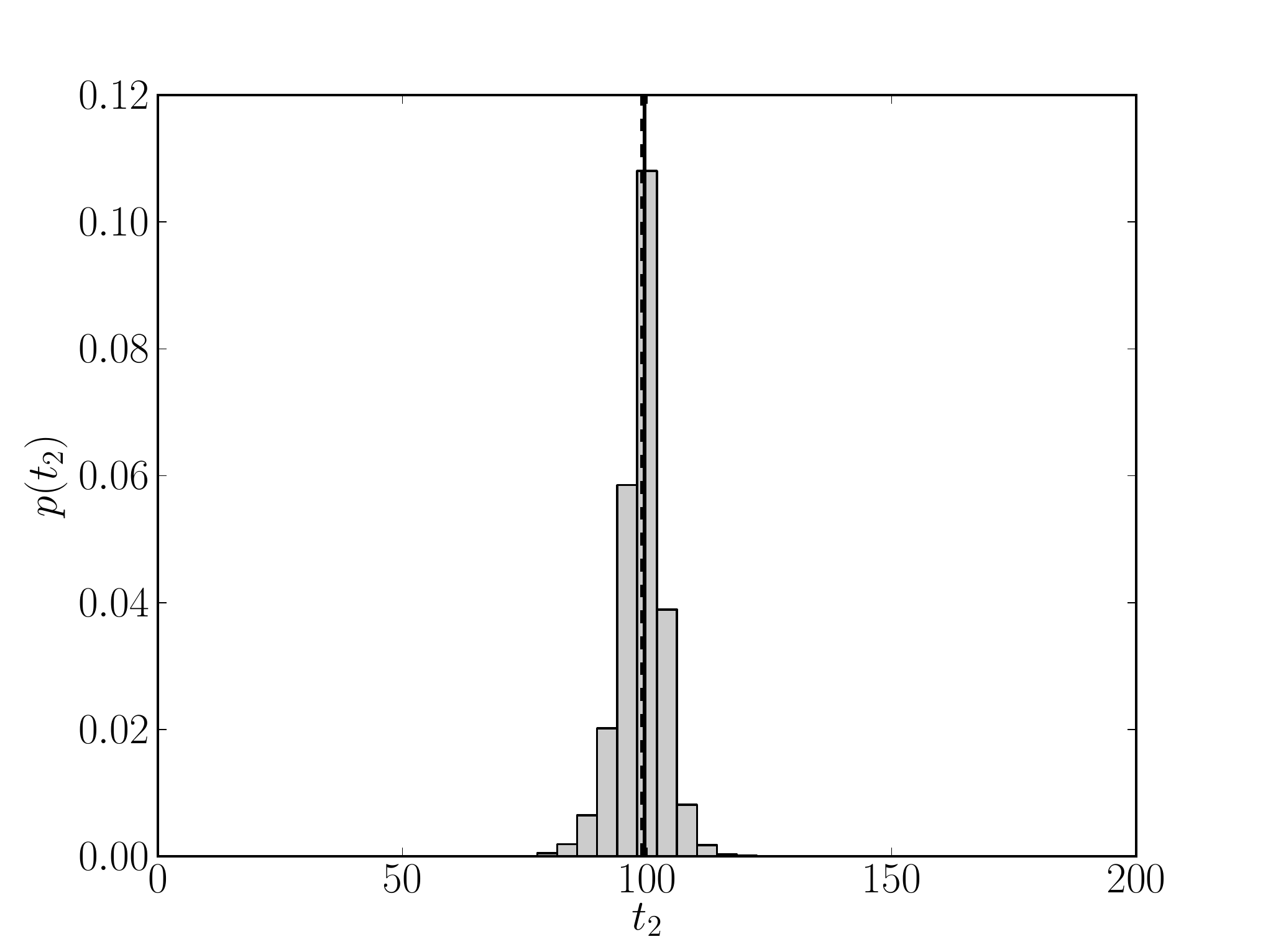}
    \label{fig:histogram_t2_pi}
  }
  \subfloat[]{
    \includegraphics[width=0.33\textwidth]{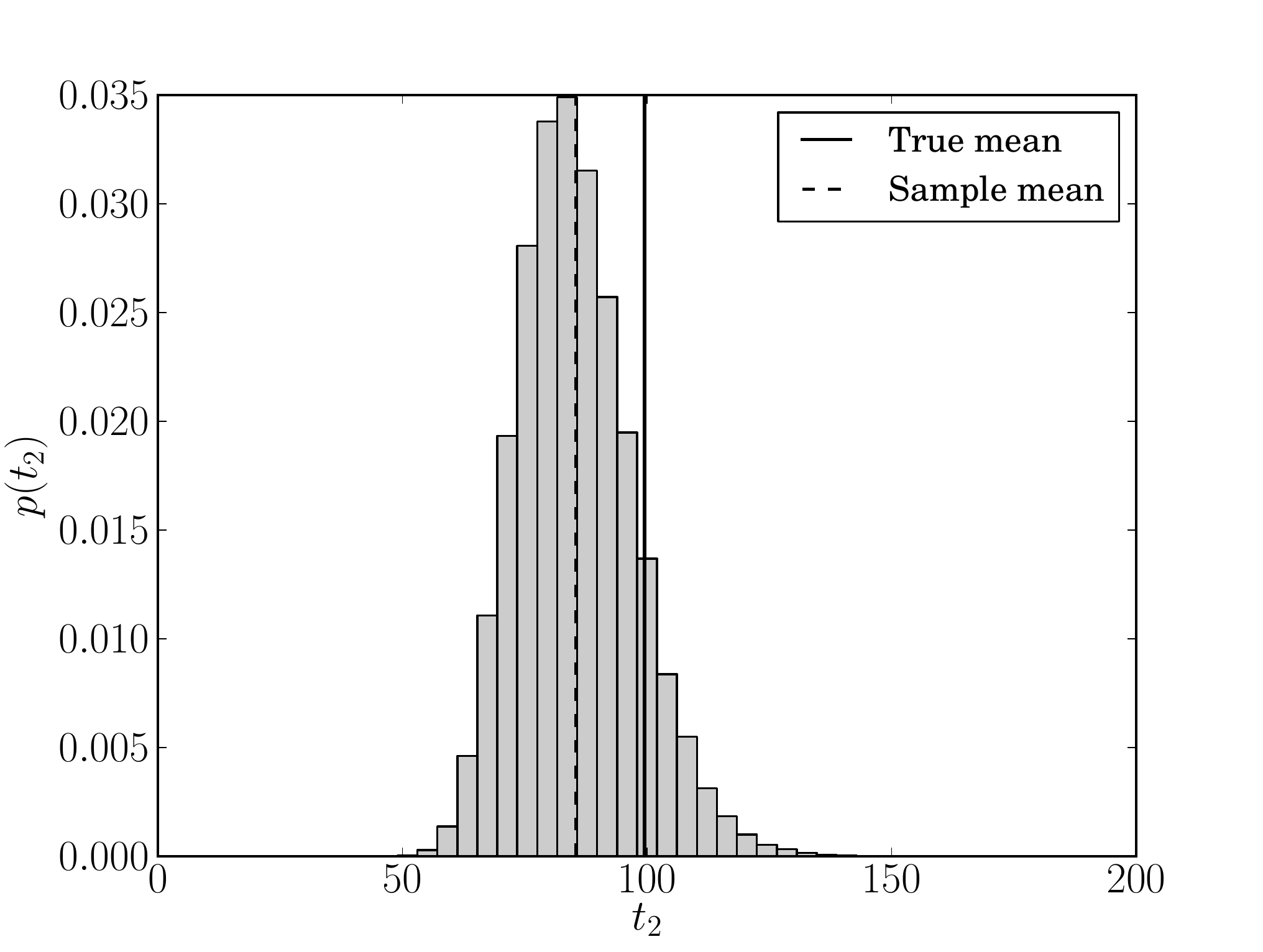}
    \label{fig:histogram_t2_nonreversible}
  }
  \subfloat[]{
    \includegraphics[width=0.33\textwidth]{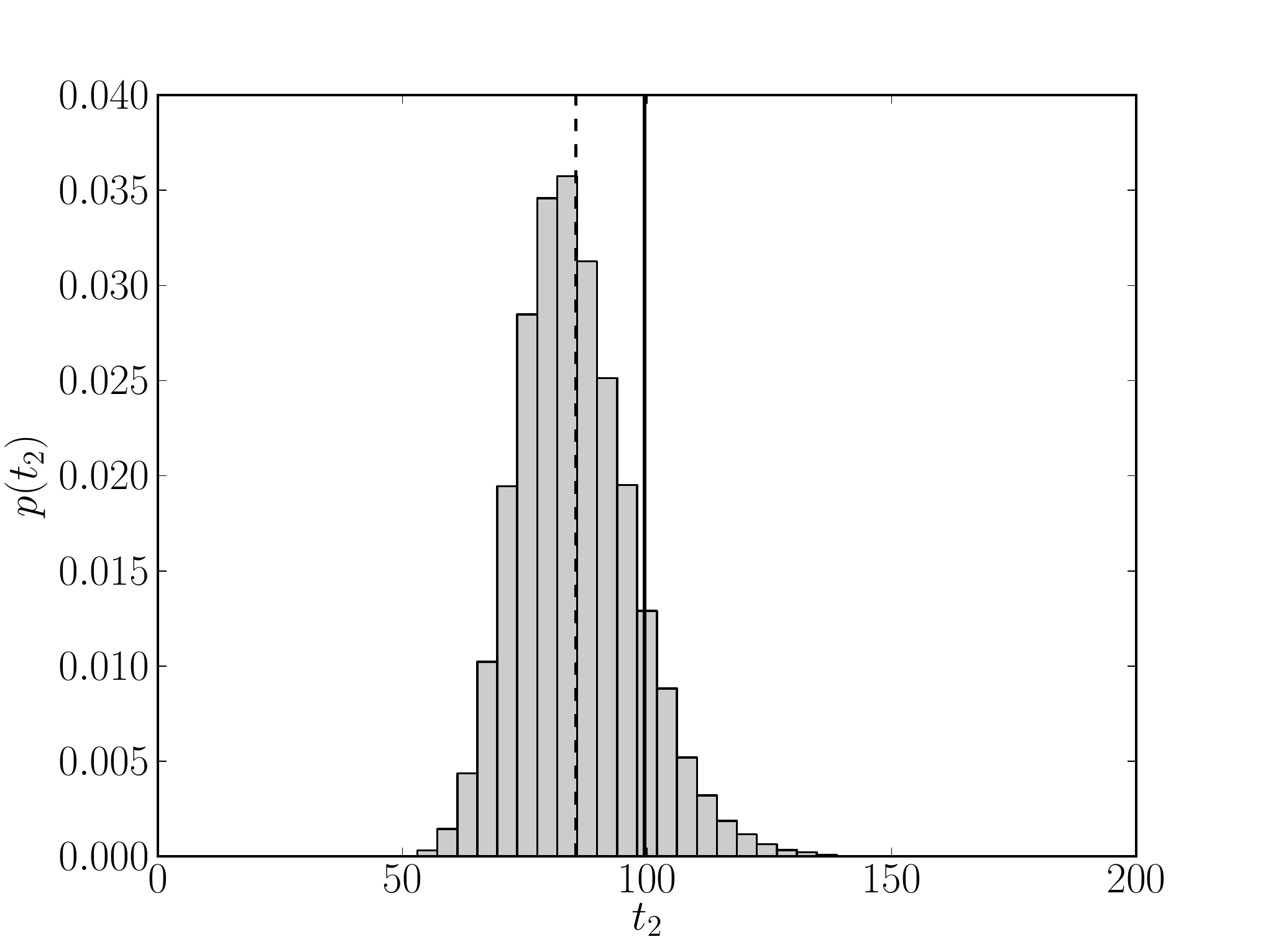}
    \label{fig:histogram_t2_reversible}
  }
\caption{Histograms for implied time scale $t_2$ corresponding to second largest eigenvalue $\lambda_2$. The histogram of timescales generated by incorporating knowledge about stationary probabilities (a) in comparison to the histogram of timescales generated by the non-reversible (b) and reversible method (c). It is clearly visible that the sample mean (dashed line) gives a more accurate prediction of the true value (solid line) due to the additional information about stationary probabilities.}
\label{fig:histogram_t2_pi_vs_reversible_vs_nonreversible}
\end{figure*}

We have used the following $3 \times 3$ transition matrix, \[T=\left(\begin{array}{ccc} 0.99 & 0.01 & 0.0 \\ 0.45 & 0.1 & 0.45 \\ 0.0 & 0.01 & 0.09\end{array}\right),\] to generate a transition counts $C$ by evolving a Markov chain for $N=5000$ steps starting from micro state $i=0$. To simulate the effect of having used an efficient enhanced sampling algorithm, the exact stationary distribution \[\mu=\left(0.4945, 0.011, 0.4945 \right)\] and the observed transition counts were used to generate a sample of $100000$ random transition matrices using the Gibbs sampling algorithm. For each of the randomly generated transition matrix the second largest eigenvalue $\lambda_2$ and the corresponding implied time scale $t_2$ was computed. For comparison, the observed transition counts were used in a similar manner to generate a sample of implied time scales without prior knowledge of the stationary distribution with and without explicitly enforcing a detailed balance condition. It is clearly visible in \autoref{fig:histogram_t2_pi_vs_reversible_vs_nonreversible} that the estimation procedure including knowledge about stationary probabilities gives a more accurate and a much sharper estimate of this quantity.

\section{Discussion and conclusion}
We have presented a Gibbs sampling algorithm for the generation of transition matrices fulfilling the detailed balance constraint with respect to a given stationary distribution. The presented algorithm shows a clear improvement in convergence speed and autocorrelation times over the algorithm presented in \cite{noe2008}. We believe that the presented algorithm will be a useful tool for Monte Carlo sampling of transition probabilities when a priori knowledge about stationary probabilities is available in addition to observed transition counts. As already pointed out in \cite{noe2008} enforcing the detailed balance condition can lead to an immense reduction in variance of certain off-diagonal elements leading to sharper estimates for kinetically relevant observables. With a priori estimates of stationary distributions available from extended ensemble simulations and an increased interest in estimating dynamical quantities of molecular systems from Markov model based approaches the outlined algorithm will hopefully become a useful statistical tool for the analysis of metastable systems.

There are several directions for future research. An improved scheme for the generation of random variates from the conditional density could further speed up the algorithm allowing the generation of transition matrices for processes with larger state spaces. Larger acceptance rates could be already achieved by finding better approximations to the optimal boundary points $x_l^{*}$, $x_u^{*}$ in the definition of the piecewise enveloping function $h(x)$. In fact the only parameter of the conditional density that is updated after a new sample is drawn is $d$. All possible values for $a,b,c$ could in principle be calculated a priori so that one might find a set of $n(n-1)/2$ optimal proposal densities each parametrized by $d$. Finding a transformation removing the parametric dependence of the conditionals on $d$ altogether seems unlikely but would of course open up the possibility for the design of even faster algorithms. 

We are currently pursuing an application of the algorithm to data sets obtained by standard molecular dynamics simulations together with estimates of the equilibrium distribution obtained from enhanced sampling algorithms such as meta-dynamics, generalized ensemble simulations or umbrella sampling to obtain sharper estimates of dynamical quantities.

\begin{acknowledgments}
The author's would like to thank two anonymous referees for helpful comments and suggestions. One of the authors would like to thank Guillermo Perez, Han Wang and Ivan Kryven for discussions and helpful suggestions. He thanks Luc Devroye for a suggestion concerning the modified rejection sampling approach. He would especially like to thank Sven Kr\"{o}nke for inspiring discussions. B. Trendelkamp-Schroer acknowledges funding by the DFG fund NO 825/3 and from the ``Center of Supramolecular Interactions'' at FU-Berlin. Frank Noe acknowledges funding from the research center Matheon.
\end{acknowledgments}

\bibliography{publication.bib}

\end{document}